\documentclass[twocolumn,aps,pre,amsmath,showpacs,floatfix]{revtex4}

\usepackage{graphicx}

\def\lessgtr{\raise2.5pt\hbox{$<$}\llap{\lower2.5pt\hbox{$>$}}}
\def\gtrless{\raise2.5pt\hbox{$>$}\llap{\lower2.5pt\hbox{$<$}}}

\begin{document}

\title{Role of structural relaxations and vibrational excitations 
in the \\
high-frequency dynamics of liquids and glasses}
\author{Song-Ho Chong}
\affiliation{Institute for Molecular Science,
Okazaki 444-8585, Japan}
\email{chong@ims.ac.jp}
\date{\today}

\begin{abstract}

We present theoretical investigation on the
high-frequency collective dynamics in liquids and glasses
at microscopic length scales and terahertz frequency region
based on the mode-coupling theory for ideal 
liquid-glass transition.
We focus on recently investigated issues from
inelastic-X-ray-scattering and computer-simulation
studies for dynamic structure
factors and longitudinal and transversal current spectra: 
the anomalous dispersion of the high-frequency sound 
velocity and the nature of the low-frequency excitation
called the boson peak.
It will be discussed how the sound mode
interferes with other low-lying modes present in the system.
Thereby, we provide a systematic 
explanation of the anomalous sound-velocity
dispersion in systems -- ranging from high temperature liquid
down to deep inside the glass state -- 
in terms of the contributions 
from the structural-relaxation processes 
and from vibrational excitations 
called the anomalous-oscillation peak (AOP).
A possibility of observing negative dispersion 
-- the {\em decrease} of the sound velocity
upon increase of the wave number -- is argued
when the sound-velocity dispersion is dominated by 
the contribution from the vibrational dynamics. 
We also show that the low-frequency excitation,
observable in both of the glass-state
longitudinal and transversal
current spectra at the same resonance frequency,
is the manifestation of the AOP.
As a consequence of the presence of the
AOP in the transversal current spectra,
it is predicted that the transversal sound velocity 
also exhibits the anomalous dispersion.
These results of the theory are demonstrated for 
a model of the Lennard-Jones system.

\end{abstract}


\maketitle

\section{Introduction}
\label{sec:introduction}

The study of high-frequency collective dynamics in liquids and glasses
at microscopic length scales and terahertz frequency region 
has been a subject of intense investigations in the past decade.
In particular,
recent development of the inelastic X-ray scattering (IXS) technique
has renewed the interest in a long-standing issue of sound 
propagation~\cite{Sette98,Ruocco01,Scopigno05}.
It is now generally accepted that a well-defined sound-like
oscillatory mode 
-- called the high-frequency or fast sound --
is supported in liquids outside 
the strict hydrodynamic region, 
down to wavelengths of a few interparticle distances,
but with a sound velocity larger than the hydrodynamic value.
Traditionally, such increase of the sound velocity upon
increase of the wave number --
called the anomalous or positive dispersion -- has 
been interpreted within the so-called viscoelastic 
model~\cite{Hansen86,Balucani94}.
In this model, relevant memory kernel in the Zwanzig-Mori or memory-function
equation for the dynamics structure factor $S_{q}(\omega)$
is modeled by an exponential function
with a single time constant $\tau$ 
reflecting structural relaxation in liquids. 
An increase of the sound velocity is predicted 
at the wave number $q$ where the condition 
$\omega_{q}^{\rm max} \tau = 1$ is fulfilled
with the resonance frequency $\omega_{q}^{\rm max}$ 
of the dynamic structure factor. 
This condition marks the transition from the low-frequency
viscous behavior to the high-frequency 
elastic behavior of the liquid,
and thus, the positive dispersion effect is ascribed to the
transition between a liquid-like to a solid-like response
of the system. 

According to the viscoelastic model, the absence of the
positive dispersion is naturally expected in glasses where the 
structural relaxation is basically frozen. 
However, molecular dynamics (MD) simulations 
have demonstrated the presence of the positive dispersion
in model glasses~\cite{Taraskin97,Ruocco00,Horbach01b,Scopigno02,Pilla04}.
A first experimental confirmation has been reported recently
from an IXS study on vitreous silica~\cite{Ruzicka04}.
The observed anomalous dispersion 
implies the presence of some
additional ``relaxation'' process which is 
active even in glasses, and stimulates an extension 
of the viscoelastic model to fully account for the
anomalies present both in liquids and glasses. 

Such extension was also motivated by recent IXS studies
on simple liquid metals near the melting 
temperature~\cite{Scopigno00,Scopigno00b,Scopigno00c,Scopigno02b,Monaco04}.
From a detailed line-shape analysis of $S_{q}(\omega)$,
the viscoelastic model was found to be unable to account for 
inelastic peaks in these systems. 
It was found, instead, that 
a two-time-scale model for the
memory kernel successfully describes
the observed spectral shapes. 
Thus, a convincing experimental demonstration is provided
for the simultaneous presence two relaxation processes
with slow and fast characteristic time scales,
termed structural (labeled $\alpha$) and
microscopic (labeled $\mu$) processes. 
Furthermore, it was pointed out that the faster
$\mu$ process is the major controller of the positive dispersion
observed in these systems since it was always found that
$\omega_{q}^{\rm max} \tau_{\alpha} > 1$ 
for the slower time scale $\tau_{\alpha}$,
whereas $\omega_{q}^{\rm max} \tau_{\mu}$ becomes close to unity
at some wave number 
for the faster time scale $\tau_{\mu}$. 
According to this experimental observation,
the positive dispersion cannot fully be ascribed to 
the structural relaxation. 

The nature of these two processes have been studied
in more detail in Ref.~\cite{Scopigno02} based on 
the MD simulation for a model of lithium.
From the analysis of the simulated $S_{q}(\omega)$
with the two-time-scale model, 
a strong temperature dependence was observed for
the time scale $\tau_{\alpha}$, corroborating that
the slower process is associated with the structural relaxation.
On the other hand, the faster $\mu$ process was found
to persist both in liquid and glass phases with almost no 
temperature dependence of its relaxation time $\tau_{\mu}$.
The persistence of the $\mu$ relaxation 
accounts for the presence of the positive 
dispersion in glasses.
Of course, a natural question would be what is relaxing 
at that low $T$, i.e.,
what is the microscopic nature of the $\mu$ dynamics.
This was studied in Ref.~\cite{Ruocco00} based on the
computer simulation for a harmonic glass. 
There, it was suggested that 
the origin of this process can be 
ascribed to the topological disorder,
i.e., the decay of the memory kernel is due to the
dephasing of different oscillatory components 
in the force fluctuations. 

Besides the peak associated with the high-frequency sound,
the MD~\cite{Horbach01b,Pilla04} 
and IXS~\cite{Scopigno03,Ruzicka04} works showed the 
presence of additional low-frequency
excitation in the dynamic structure factors
$S_{q}(\omega)$ or in the related longitudinal current
spectra $\propto \omega^{2} S_{q}(\omega)$.
This second low-frequency excitation -- also called the boson peak --
exhibits some general characteristics:
it appears in the spectra at $q$ values larger than
some fraction of the structure-factor-peak position,
and its resonance position is weakly $q$-dependent.
It was conjectured that 
the low-frequency excitation reflects the 
transversal dynamics, and
the appearance of the transversal mode
in the longitudinal current 
spectra was interpreted as being due to ``mixing'' 
phenomena caused by microscopic disorder~\cite{Sampoli97}. 
The main evidence supporting this conjecture lies in the 
simulation results that the low-frequency excitation
appears at the same resonance frequency
in both of the transversal and longitudinal
current spectra, but its intensity is more enhanced in the former. 
Another support was inferred 
from studies on the density dependence. 
The computational work in Ref.~\cite{Pilla04} demonstrated 
that the resonance frequency of the low-frequency
excitation in the transversal current spectra
increases with increasing density. 
Experimentally, it is observed that upon densification
the boson-peak energy shifts to higher energy and its intensity 
strongly decreases~\cite{Sugai96,Inamura98,Inamura99,Inamura00}.
This parallel of the density dependence was claimed to also suggest
that the boson-peak-like low-frequency excitation arises 
from the transversal dynamics. 

In this paper, we present a microscopic and unified 
explanation of the mentioned features of the
high-frequency dynamics in liquids and glasses
-- the anomalous dispersion of the high-frequency sound 
velocity and the nature of the low-frequency excitation -- 
which have so far been investigated somewhat independently.  
This will be done based on the mode-coupling theory (MCT) for 
ideal liquid-glass transition~\cite{Goetze91b}.
Originally, the MCT was developed to deal with 
structural-relaxation processes 
which evolve as precursor of the glass transition.
On the other hand, the glass transition also modifies the
short-time or the high-frequency dynamics.
The study of these modifications was the main subject in 
Ref.~\cite{Goetze00}.
It was shown there that the strong interaction 
between density fluctuations
at microscopic length scales and the arrested glass 
structure causes
an anomalous-oscillation peak (AOP), which exhibits the
properties of the boson peak.
As will be demonstrated in the present study,
the AOP persists in liquid states as well as in glass states.
It will be discussed how the sound mode
interferes with other low-lying modes present in the system.
Thereby, we provide a systematic 
explanation of the anomalous sound-velocity
dispersion in systems -- ranging from high temperature liquid
down to deep inside the glass state -- 
in terms of the contributions 
from the structural-relaxation processes 
and from the vibrational modes building the AOP.
A possibility of observing negative dispersion 
-- the {\em decrease} of the sound velocity
upon increase of the wave number -- is argued
when the sound-velocity dispersion is dominated by 
the contribution from the vibrational dynamics. 
We also show that the low-frequency excitation,
observable in both of the glass-state
longitudinal and transversal
current spectra at the same resonance frequency,
is the manifestation of the AOP.
This seems to be an alternative interpretation to 
the mixing phenomena mentioned above. 
As a natural consequence of the presence of the
AOP in the transversal current spectra,
it is predicted that the transversal sound velocity 
also exhibits the anomalous dispersion.

The paper is organized as follows.
In Sec.~\ref{sec:theory},
the basic MCT equations of motion are formulated, and 
the model for the demonstration of our theoretical
results is specified.
Section~\ref{sec:sound} deals with the anomalous
sound-velocity dispersion of the high-frequency sound,
while in Sec.~\ref{sec:transversal} we discuss
features of glass-state
longitudinal and transversal current 
spectra at low frequencies. 
Section~\ref{sec:conclusions} concludes the paper.

\section{Basic theory}
\label{sec:theory}

\subsection{MCT equations of motion}
\label{sec:theory-1}

The basic quantity describing the equilibrium structure
of a simple system is the static structure factor
$S_{q} = \langle | \, \rho_{\vec q} \, |^{2} \rangle$
defined in terms of the density fluctuations 
for wave vector ${\vec q}$,
$\rho_{\vec q} = \sum_{i=1}^{N} 
\exp(i {\vec q} \cdot {\vec r}_{i} \,) / \sqrt{N}$.
Here, $N$ is the total number of particles in the system
distributed with the average density $\rho$, 
and ${\vec r}_{i}$ denotes the position of the $i$th particle.
For homogeneous and isotropic system, 
which we assume throughout the paper, 
the static structure factor 
depends only on the modulus $q = | \, {\vec q} \, |$.
The most relevant variables characterizing the structural
changes as a function of time $t$ are the
density correlators
$\phi_{q}(t) = \langle \rho_{\vec q}(t)^{*} \rho_{\vec q}(0) \rangle
\, / \, S_{q}$
which are normalized to unity at $t = 0$. 
The short-time asymptote of these functions is given by
$\phi_{q}(t) = 1 - (1/2) \Omega_{q}^{2} t^{2} + \cdots$,
where $\Omega_{q}^{2}$ denotes the square of the characteristic
frequency $\Omega_{q}^{2} = q^{2} v^{2} / S_{q}$~\cite{Balucani94}.
Here $v = \sqrt{k_{\rm B}T/M}$ with Boltzmann's constant $k_{\rm B}$
denotes the thermal 
velocity of particle of mass $M$ at temperature $T$.
In the small wave-vector limit, one obtains
$\Omega_{q} = v_{0} q + O(q^{3})$ with 
the isothermal sound velocity $v_{0} = v / \sqrt{S_{q=0}}$.
Within the Zwanzig-Mori formalism~\cite{Balucani94}
one can derive the following exact equation of motion
\begin{subequations}
\label{eq:GLE-phi}
\begin{equation}
\partial_{t}^{2} \phi_{q}(t) + \Omega_{q}^{2} \, \phi_{q}(t) +
\Omega_{q}^{2} \int_{0}^{t} dt^{\prime} \,
m_{q}(t-t^{\prime}) \,
\partial_{t^{\prime}} \phi_{q}(t^{\prime}) = 0,
\label{eq:GLE-phi-a}
\end{equation}
in which the relaxation kernel $m_{q}(t)$ is a correlation function of
fluctuating forces.
Let us introduce Fourier-Laplace transformations to map
$\phi_{q}$, $m_{q}$, and similar functions from the time
domain onto the frequency domain according to the convention
$\phi_{q}(\omega) = i \int_{0}^{\infty} dt \, 
\exp(i \omega t) \phi_{q}(t) = \phi_{q}'(\omega) +
i \phi_{q}''(\omega)$.
Equation~(\ref{eq:GLE-phi-a}) is equivalent to the representation
\begin{equation}
\phi_{q}(\omega) = - 1 \, / \, 
\{\, \omega - \Omega_{q}^{2} \, / \, 
[\, \omega + \Omega_{q}^{2} m_{q}(\omega) \, ] \, \}.
\label{eq:GLE-phi-b}
\end{equation}
\end{subequations}

Within the MCT, the kernel $m_{q}(t)$ is given by a mode-coupling
functional of density correlators 
describing the cage effect of dense systems,
\begin{subequations}
\label{eq:MCT-phi}
\begin{equation}
m_{q}(t) = {\cal F}_{q}[\phi(t)].
\label{eq:MCT-phi-a}
\end{equation}
Here the mode-coupling functional reads
${\cal F}_{q}[\tilde{f}] = 1/(2\pi)^{3} \int d{\vec k} \,
V({\vec q} \, ; {\vec k}, {\vec p} \, ) \,
\tilde{f}_{k} \tilde{f}_{p}$
with ${\vec p} = {\vec q} - {\vec k}$.
The coupling coefficients 
$V({\vec q} \, ; {\vec k}, {\vec p} \, )$ are
determined by the equilibrium structure.
Within the convolution approximation for triple correlations,
they are given by
$V({\vec q} \, ; {\vec k}, {\vec p} \, ) =
\rho S_{q} S_{k} S_{p} \,
[ \, {\vec q} \cdot ({\vec k} c_{k} + {\vec p} c_{p}) \,]^{2} \, / \,
(2 q^{4})$, where
$c_{q}$ denotes the direct correlation function
related to $S_{q}$ via the Ornstein-Zernike equation,
$\rho c_{q} = 1 - 1/S_{q}$~\cite{Hansen86}.
For details, the reader is referred to Ref.~\cite{Goetze91b}.
In the following, wave-vector integrals will be approximated
by Riemann sums. 
The mode-coupling functional can then be written as a 
quadratic polynomial
\begin{equation}
{\cal F}_{q}[\tilde{f}] = \sum_{k,p} V_{q,kp} \,
\tilde{f}_{k} \tilde{f}_{p},
\label{eq:MCT-phi-b}
\end{equation}
where the indices run over the 
discretized wave-number grids. 
The coefficients $V_{q,kp}$ are trivially related to 
$V({\vec q} \, ; {\vec k}, {\vec p} \,)$.
The $q \to 0$ limit of the functional reads
\begin{equation}
{\cal F}_{0}[\tilde{f}] = \sum_{k} V_{k} \tilde{f}_{k}^{2},
\label{eq:MCT-phi-c}
\end{equation}
\end{subequations}
with 
$V_{k} = 
(\rho S_{0}/4 \pi^{2}) 
k^{4} S_{k}^{2}
[ \, c_{k}^{2} + 2 (kc_{k}^{\prime}) c_{k} / 3 +
(kc_{k}^{\prime})^{2} / 5 \,]$.

Equations~(\ref{eq:GLE-phi}) and (\ref{eq:MCT-phi}) 
are closed, provided equilibrium 
static quantities are known as input.
In the following, the results will be demonstrated for the
Lennard-Jones (LJ) system:
the interaction between two particles is given by
the LJ interaction
$V(r) = 4 \epsilon_{\rm LJ} 
\{ (\sigma_{\rm LJ}/r)^{12} - (\sigma_{\rm LJ}/r)^{6} \}$.
The first MCT work for the LJ system has been done
in Refs.~\cite{Bengtzelius86,Bengtzelius86b}
with the optimized random-phase
approximation for $S_{q}$~\cite{Hansen86}.
In the present study, 
$S_{q}$ is evaluated within the
Percus-Yevick approximation~\cite{Hansen86}.
Wave numbers will be considered up to a cutoff value 
$q^{*} = 80/\sigma_{\rm LJ}$,
and they are discretized to $400$ grid points.
From here on, all quantities are expressed in reduced units
with the unit of length $\sigma_{\rm LJ}$, the unit of energy 
$\epsilon_{\rm LJ}$ (setting $k_{\rm B} = 1$),
and the unit of time $(M\sigma_{\rm LJ}^{2}/\epsilon_{\rm LJ})^{1/2}$.
The dynamics by varying $T$ shall be considered for 
a fixed density $\rho = 1.093$, except for 
Sec.~\ref{sec:transversal-2} where
the dynamics at a lower density $\rho = 0.95$ is discussed. 
(We notice that the triple point of the LJ system is
at $\rho^{*} \approx 0.85$ and $T^{*} \approx 0.68$~\cite{Hansen69}.)

\subsection{Ideal glass states}
\label{sec:theory-2}

\begin{figure}[htb]
\includegraphics[width=0.8\linewidth]{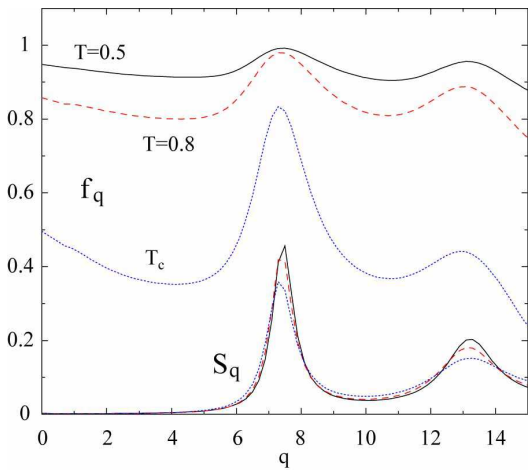}
\caption{
Debye-Waller factor $f_{q}$ and one-tenth of the 
static structure factor $S_{q}$ at $\rho = 1.093$ for
$T = 0.5$ (solid lines), $T = 0.8$ (dashed lines),
and $T = T_{\rm c} \approx 1.637$ (dotted lines).}
\label{fig:fq-Sq}
\end{figure}

The specified model exhibits a fold bifurcation~\cite{Goetze91b}.
For small coupling constants $V_{q,kp}$,
the correlators $\phi_{q}(t)$ and the kernels $m_{q}(t)$
decay to zero for long times, 
$\phi_{q}(t \to \infty) = 0$ and $m_{q}(t \to \infty) = 0$.
In this case, the spectra $\phi_{q}''(\omega)$ and
$m_{q}''(\omega)$ are continuous in $\omega$.  
In particular, there hold
$\lim_{\omega \to 0} \omega \phi_{q}(\omega) = 0$
and $\lim_{\omega \to 0} \omega m_{q}(\omega) = 0$.
Density fluctuations created at time $t=0$
disappear for long times and the same holds for the
force fluctuations as expected for an ergodic liquid.
For large coupling constants, on the other hand, there is arrest 
of density fluctuations for long times:
$\phi_{q}(t \to \infty) = f_{q}$, $0<f_{q}<1$.
Thus, nonergodic dynamics is obtained in which 
the perturbed system does not return to the equilibrium
state.
Similarly, there is arrest of the force fluctuations, 
$m_{q}(t \to \infty) = C_{q} > 0$.
The nonergodicity parameters 
$f_{q}$ and $C_{q}$ are connected via~\cite{Bengtzelius84}
\begin{equation}
f_{q} = C_{q} \, / \, (1 + C_{q}), \quad
C_{q} = {\cal F}_{q}[f].
\label{eq:DW}
\end{equation}
For this strong-coupling solution, the kernel exhibits a 
zero-frequency pole,
$\lim_{\omega \to 0} \omega m_{q}(\omega) = - C_{q}$. 
The fluctuation spectrum $\phi_{q}''(\omega) \propto S_{q}(\omega)$
exhibits a strictly elastic peak:
$\phi_{q}''(\omega) = \pi f_{q} \delta(\omega) + 
\mbox{regular terms}$.
This is the signature for a solid with $f_{q}$ denoting
its Debye-Waller factor.
Hence, the strong coupling solution deals with a disordered
solid; it is a model for an ideal glass state.
If one increases the coupling constants smoothly 
from small to large values, one finds a singular change of the
solution from the ergodic liquid to the nonergodic glass state,
i.e., an idealized liquid-glass transition.
For simple-liquid models, the transition occurs upon cooling
at some critical temperature $T_{\rm c}$ or upon compression
at some critical density $\rho_{\rm c}$~\cite{Bengtzelius84}.
For the LJ model under study where the temperature is
considered as a control parameter, 
one finds $T_{\rm c} \approx 1.637$ for $\rho = 1.093$
within the Percus-Yevick approximation for $S_{q}$. 
If $T$ decreases below $T_{\rm c}$,
$f_{q}$ increases above its value at
the critical point, called the critical nonergodicity parameter
or the plateau $f_{q}^{\rm c}$,
as demonstrated in Fig.~\ref{fig:fq-Sq}.

\subsection{MCT liquid-glass-transition dynamics}
\label{sec:theory-3}

As precursor of the ideal liquid-glass transition,
MCT predicts the evolution of the glassy dynamics
which are stretched over many decades.
Such MCT scenario 
has been comprehensively discussed for the hard-sphere system
in Refs.~\cite{Franosch97,Fuchs98}.
This subsection compiles the MCT universal predictions
for the dynamics near the liquid-glass transition to an extent which
is necessary for understanding the present article.

We start from introducing some concepts to 
describe the MCT-liquid-glass-transition dynamics~\cite{Goetze91b}.
In the space of control
parameters, a smooth function $\sigma$ is defined near the
transition point, called the separation parameter. 
Glass states are characterized by $\sigma > 0$, 
liquid states by $\sigma < 0$,
and $\sigma = 0$ defines the transition point. 
When only the temperature $T$ is considered as a control parameter
near the transition point, one can write for small
distance parameters $\epsilon = (T_{\rm c} - T) / T_{\rm c}$:
$\sigma = C_{T} \epsilon$, $C_{T} > 0$. 
In addition, the transition point is
characterized by a time scale $t_0$ and by a number $\lambda$, 
$0 < \lambda < 1$.
The scale $t_0$ specifies properties of the short-time
transient dynamics,
and $\lambda$ is called the exponent parameter. 
The latter determines a certain
number $B > 0$, the critical exponent $a, \, 0 < a \leq 1/2$, and
the von-Schweidler exponent $b, \, 0 < b \leq 1$.
For the specified LJ model at $\rho = 1.093$, we have
$C_{T} = 0.226$, $\lambda = 0.697$, $a = 0.328$, 
$b = 0.646$, $B = 0.669$, and $t_{0} = 0.0125$. 

Let us consider the correlator 
$\phi_{X}(t) = \langle X(t)^{*} X(0) \rangle$ 
of some variable $X$ coupling to the density fluctuations.
Its nonergodicity parameter $f_{X} = \phi_{X}(t \to \infty)$
exhibits a square-root singularity near the transition,
\begin{equation}
f_{X} = f_{X}^{\rm c} + h_{X} \sqrt{\sigma/(1-\lambda)}, \quad
\sigma > 0, \quad \sigma \to 0.
\label{eq:plateau-X}
\end{equation}
The critical nonergodicity parameter $f_{X}^{\rm c}$
and the critical amplitude $h_{X}$ are equilibrium
quantities to be calculated from the relevant mode-coupling
functionals at the critical point $T_{\rm c}$.
At the transition, the correlator exhibits a 
power-law decay that is specified by the critical exponent $a$.
In a leading-order expansion in $(1/t)^{a}$, one gets
\begin{equation}
\phi_{X}(t) = f_{X}^{\rm c} + h_{X} (t_{0}/t)^{a}, \quad
\sigma = 0, \quad (t/t_{0}) \to \infty.
\label{eq:critical-X}
\end{equation}

For small values of $\epsilon$, 
there is a large time interval, where $\phi_{X}(t)$ 
is close to its critical nonergodicity parameter 
$f_{X}^{\rm c}$.
Solving the equations of motion asymptotically
for this plateau regime, one gets in leading order in the
small quantities $\phi_{X}(t) - f_{X}^{\rm c}$ 
the factorization theorem:
\begin{subequations}
\label{eq:beta-correlator}
\begin{equation}
\phi_{X}(t) = f_{X}^{\rm c} + h_{X} G(t).
\label{eq:beta-correlator-a}
\end{equation}
The function $G(t)$, called the $\beta$ correlator,
is the same for all variables $X$, and 
describes the complete dependence on time and control
parameter via the first scaling law
\begin{equation}
G(t) = \sqrt{ | \, \sigma \, |} \, g_{\pm}(t/t_{\sigma}),
\quad \sigma \gtrless 0.
\label{eq:beta-correlator-b}
\end{equation}
\end{subequations}
Here, $t_{\sigma} = t_{0} / | \, \sigma \, |^{(1/2a)}$ 
is the first critical time scale.
The master functions $g_{\pm}(\hat{t})$ are determined by 
$\lambda$~\cite{Goetze90}. 
For $\hat{t} \ll 1$, the master functions are the same on 
both sides of the transition point,
and one gets the power law
$g_{\pm}(\hat{t} \to 0) = 1 / \hat{t}^{a}$ so that
Eq.~(\ref{eq:critical-X}) is reproduced for fixed large $t$
and $\sigma$ tending to zero.
For $\hat{t} \gg 1$, one gets for glasses
$g_{+}(\hat{t} \gg 1) = 1/\sqrt{1-\lambda}$,
so that Eq.~(\ref{eq:plateau-X}) is reproduced. 
On the other hand, for liquids,
one finds 
for the large rescaled time $\hat{t}$:
$g_{-}(\hat{t} \to \infty) = - B \hat{t}^{b} + O(1/\hat{t}^{b})$.
Substituting this result into 
Eqs.~(\ref{eq:beta-correlator}), one obtains
von Schweidler's law for the decay of the liquid correlator below the
plateau $f_{X}^{\rm c}$:
\begin{equation}
\phi_{X}(t) = f_{X}^{\rm c} - h_{X} (t/t_{\sigma}^{\prime})^{b}, \quad
t_{\sigma} \ll t, \quad
\sigma \to -0.
\label{eq:phi-X-von}
\end{equation}
The control-parameter dependence is given via
the second critical time scale 
$t_{\sigma}^{\prime} =
t_{0} \, B^{-1/b} \, / \, |\sigma|^{\gamma}$
with $\gamma = (1/2a) + (1/2b)$. 
The dynamical process described by the cited results is called
the $\beta$-process.
The $\beta$ correlator $G(t)$ describes in leading order the decay
of the correlator towards the plateau $f_{X}^{\rm c}$ within
the interval $t_{0} \ll t \ll t_{\sigma}$.
For $t \gg t_{\sigma}$, 
the glass correlator arrests at $f_{X}$, while 
the liquid correlator exhibits von Schweidler's law.

The decay of the liquid correlator
below the plateau $f_{X}^{\rm c}$ is called the 
$\alpha$-process. 
For this process, there holds in leading order for $\sigma \to -0$
the second scaling law, 
$\phi_{X}(t) = \tilde{\phi}_{X}(\tilde{t})$ with 
$\tilde{t} = t / t_{\sigma}^{\prime}$,
which is also referred to as the superposition principle. 
The control-parameter-independent shape function 
$\tilde{\phi}_{X}(\tilde{t})$ is to be evaluated from the mode-coupling
functionals at the critical point. 
For short rescaled times $\tilde{t}$,
one gets 
$\tilde{\phi}_{X}(\tilde{t}) = 
f_{X}^{\rm c} - h_{X} \tilde{t}^{b} + O(\tilde{t}^{2b})$, 
so that Eq.~(\ref{eq:phi-X-von}) is reproduced.
The ranges of applicability of the first and the second scaling laws
overlap;
both scaling laws yield von Schweidler's law for 
$t_{\sigma} \ll t \ll t_{\sigma}^{\prime}$.

\subsection{Description of the glass-state dynamics}
\label{sec:theory-4}

For later convenience, we introduce a reformulation of the MCT
equations of motion which is more appropriate 
in handling the dynamics in glass states~\cite{Goetze91b}. 
We map the density correlators
$\phi_{q}$ to new ones $\hat{\phi}_{q}$ by
\begin{subequations}
\begin{equation}
\phi_{q}(t) = f_{q} + (1 - f_{q}) \, \hat{\phi}_{q}(t).
\label{eq:GLE-hat-phi-a}
\end{equation}
This amounts to dealing only with the decay relative to the frozen
amorphous structure described by $f_{q}$. 
Introducing new characteristic frequencies
$\hat{\Omega}_{q}$ by
\begin{equation}
\hat{\Omega}_{q}^{2} = \Omega_{q}^{2} \, / \, (1-f_{q}),
\label{eq:GLE-hat-phi-b}
\end{equation}
one obtains for short times 
$\hat{\phi}_{q}(t) = 1 - (1/2) \hat{\Omega}_{q}^{2} t^{2} + \cdots$
in analogy to the one for $\phi_{q}(t)$.
In the small wave-vector limit, there holds
$\hat{\Omega}_{q} = \hat{v}_{0} q + O(q^{3})$
with $\hat{v}_{0} = v_{0} / \sqrt{1-f_{0}}$.
The modification of the sound velocity reflects the
reduced compressibility in nonergodic glass states. 
Substitution of these results into Eq.~(\ref{eq:GLE-phi-a}) 
reproduces the MCT
equations of motion with $\phi_{q}$, $\Omega_{q}$, and $m_{q}$
replaced by $\hat{\phi}_{q}$, $\hat{\Omega}_{q}$,
and $\hat{m}_{q}$, respectively.
Here, the new relaxation kernel is related to the original one by
\begin{equation}
m_{q}(t) = C_{q} + (1+C_{q}) \, \hat{m}_{q}(t).
\label{eq:GLE-hat-phi-c}
\end{equation}
The new correlator has a vanishing long-time limit,
$\lim_{t \to \infty} \hat{\phi}_{q}(t) = 0$, and
correspondingly the Fourier-Laplace transform exhibits 
a regular zero-frequency behavior 
$\lim_{\omega \to 0} \omega \hat{\phi}_{q}(\omega) = 0$.
Combining Eqs.~(\ref{eq:DW}) and (\ref{eq:GLE-hat-phi-c}), 
one also concludes that
$\lim_{t \to \infty} \hat{m}_{q}(t) = 0$ and 
$\lim_{\omega \to 0} \omega \hat{m}_{q}(\omega) = 0$.
Since the equation of motion does not change its form, 
one gets as the analog of Eq.~(\ref{eq:GLE-phi-b})
\begin{equation}
\hat{\phi}_{q}(\omega) = - 1 \, / \, 
\{\, \omega - \hat{\Omega}_{q}^{2} \, / \, 
[\, \omega + \hat{\Omega}_{q}^{2} \hat{m}_{q}(\omega) \, ] \, \}.
\label{eq:GLE-hat-phi-d}
\end{equation}
\end{subequations}

The reformulated equation of motion for $\hat{\phi}_{q}(t)$
can be closed by finding an expression for the new kernel
$\hat{m}_{q}(t)$ as a new mode-coupling functional
$\hat{\cal F}_{q}$. 
One finds combining Eqs.~(\ref{eq:GLE-hat-phi-a}) and 
(\ref{eq:GLE-hat-phi-c}) with
Eqs.~(\ref{eq:MCT-phi-a}) and (\ref{eq:MCT-phi-b})
that $\hat{\cal F}_{q}$ is given by  
a sum of linear and quadratic terms
\begin{subequations}
\label{eq:MCT-hat-phi}
\begin{equation}
\hat{m}_{q}(t) = \hat{\cal F}[\hat{\phi}(t)] =
\hat{\cal F}^{(1)}[\hat{\phi}(t)] +
\hat{\cal F}^{(2)}[\hat{\phi}(t)],
\label{eq:MCT-hat-phi-a}
\end{equation}
where
\begin{equation}
\hat{\cal F}_{q}^{(1)}[\tilde{f}] =
\sum_{k} \hat{V}_{q,k} \tilde{f}_{k}, \quad
\hat{\cal F}_{q}^{(2)}[\tilde{f}] =
\sum_{k,p} \hat{V}_{q,kp} \tilde{f}_{k} \tilde{f}_{p}, 
\end{equation}
with renormalized coefficients
\begin{eqnarray}
\hat{V}_{q,k} &=& 2 ( 1-f_{q} ) \sum_{p} V_{q,kp} (1-f_{k}) f_{p},
\label{eq:MCT-hat-phi-b}
\\
\hat{V}_{q,kp} &=& ( 1-f_{q} ) V_{q,kp} (1-f_{k}) (1-f_{p}).
\label{eq:MCT-hat-phi-c}
\end{eqnarray}
\end{subequations}
Correspondingly, one finds from Eq.~(\ref{eq:MCT-phi-c})
for the zero wave-number limit of the kernel
$m(t) = m_{q = 0}(t)$
\begin{subequations}
\label{eq:MCT-hat-phi-qnull}
\begin{equation}
\hat{m}(t) = \hat{m}^{(1)}(t) + \hat{m}^{(2)}(t), 
\label{eq:MCT-hat-phi-d}
\end{equation}
in which linear and quadratic terms are given by
\begin{equation}
\hat{m}^{(1)}(t) = \sum_{k} 
\hat{V}_{k}^{(1)} \, \hat{\phi}_{k}(t), \quad
\hat{m}^{(2)}(t) = \sum_{k} 
\hat{V}_{k}^{(2)} \, \hat{\phi}_{k}(t)^{2}, 
\label{eq:MCT-hat-phi-qnull-b}
\end{equation}
with new coefficients
\begin{eqnarray}
\hat{V}_{k}^{(1)} &=& 2 (1-f_{0}) V_{k} f_{k} ( 1-f_{k}),
\label{eq:MCT-hat-phi-e}
\\
\hat{V}_{k}^{(2)} &=& (1-f_{0}) V_{k} ( 1-f_{k})^{2}.
\label{eq:MCT-hat-phi-f}
\end{eqnarray}
\end{subequations}
As a result, equations of motion are produced, which are of the same
form as the original MCT 
equations~(\ref{eq:GLE-phi}) and (\ref{eq:MCT-phi}).
But in addition to the quadratic mode-coupling term,
there appears a linear term, and the original mode-coupling
coefficients are renormalized to hatted ones. 
The linear term describes interactions of density fluctuations
with the arrested amorphous structure, while the
quadratic one deals with two-mode decay processes.

\subsection{Longitudinal and transversal current correlators}
\label{sec:theory-5}

Current correlators are also relevant variables 
describing the dynamics of dense systems, in particular,
when one is interested in vibrational properties. 
These are defined in terms of the 
current fluctuations,
$j_{\vec q}^{\alpha} = \sum_{i=1}^{N} v_{i}^{\alpha} 
\exp(i {\vec q} \cdot {\vec r}_{i} \,) / \sqrt{N}$,
where $\alpha$ ($=x$, $y$, or $z$) refers to the spatial component
and $v_{i}^{\alpha}$ denotes the velocity of the $i$th particle.
Since $\langle v_{i}^{\alpha} v_{j}^{\beta} \rangle = 
\delta_{ij} \delta_{\alpha \beta} \, v^{2}$,
one gets the static correlation
$\langle j_{\vec q}^{\alpha \, *} j_{\vec q}^{\beta} \rangle =
\delta_{\alpha \beta} \, v^{2}$.
The current correlators for isotropic system
can be represented by two independent functions
which depend only on the modulus $q$:
\begin{equation}
\langle j_{\vec q}^{\alpha}(t)^{*} j_{\vec q}^{\beta}(0) \rangle / v^{2} =
\hat{q}^{\alpha} \hat{q}^{\beta} \,
\phi_{q}^{\rm L}(t) +
[ \delta_{\alpha \beta} - \hat{q}^{\alpha} \hat{q}^{\beta} ] \, 
\phi_{q}^{\rm T}(t).
\label{eq:L-T-def}
\end{equation}
Here $\hat{q}^{\alpha}$ denotes 
the $\alpha$ component of the unit vector along
${\vec q}$. 
The functions $\phi_{q}^{\rm L}(t)$ and $\phi_{q}^{\rm T}(t)$ are 
called the longitudinal and transversal current correlators, 
respectively, and they are normalized to unity at $t=0$.

\subsubsection{Longitudinal current correlator}

The density fluctuations and the longitudinal current fluctuations 
are related via the continuity equation, 
$\partial_{t} \rho_{\vec q} = i {\vec q} \cdot {\vec j}_{\vec q}$,
and there holds
$\partial_{t}^{2} \phi_{q}(t) = - \Omega_{q}^{2} \phi_{q}^{\rm L}(t)$.
Their Fourier-Laplace transforms therefore satisfy
$\omega[1+\omega \phi_{q}(\omega)] = 
\Omega_{q}^{2} \phi_{q}^{\rm L}(\omega)$,
and one obtains from Eq.~(\ref{eq:GLE-phi-b})
for the 
longitudinal current spectrum $\phi_{q}^{{\rm L}''}(\omega)$
\begin{equation}
\phi_{q}^{{\rm L}''}(\omega) =
\frac{\omega^{2} \, \Omega_{q}^{2} \, m_{q}''(\omega)}
{[\omega^{2} - \Omega_{q}^{2} \, \Delta_{q}(\omega)]^{2} +
[\omega \, \Omega_{q}^{2} \, m_{q}''(\omega)]^{2}}, 
\label{eq:GLE-L-b}
\end{equation}
with $\Delta_{q}(\omega) = 1 - \omega \, m_{q}'(\omega)$. 

The continuity equation holds also in glass states,
$\partial_{t}^{2} \hat{\phi}_{q}(t) = -
\hat{\Omega}_{q}^{2} \, \phi_{q}^{\rm L}(t)$
in terms hatted variables defined in 
Eqs.~(\ref{eq:GLE-hat-phi-a}) and (\ref{eq:GLE-hat-phi-b}),
and their spectra are related via 
\begin{equation}
\omega^{2} \hat{\phi}_{q}''(\omega) = 
\hat{\Omega}_{q}^{2} \phi_{q}^{{\rm L}''}(\omega).
\label{eq:continuity}
\end{equation}
We therefore have the same expression for 
$\phi_{q}^{{\rm L}''}(\omega)$ for glass states as in 
Eq.~(\ref{eq:GLE-L-b}) but with all quantities 
in the right-hand side replaced by hatted ones:
\begin{equation}
\phi_{q}^{{\rm L}''}(\omega) =
\frac{\omega^{2} \, \hat{\Omega}_{q}^{2} \, \hat{m}_{q}''(\omega)}
{[\omega^{2} - \hat{\Omega}_{q}^{2} \, \hat{\Delta}_{q}(\omega)]^{2} +
[\omega \, \hat{\Omega}_{q}^{2} \, \hat{m}_{q}''(\omega)]^{2}}
\quad \mbox{(glass)},
\label{eq:GLE-L-glass}
\end{equation}
with $\hat{\Delta}_{q} = 1 - \omega \hat{m}_{q}'(\omega)$.
Thus, the same relaxation kernel $m_{q}$ or $\hat{m}_{q}$
as the one for the density correlator
describes the dynamics of the longitudinal current.

\subsubsection{Transversal current correlator}

The MCT equations of motion for the transversal current
correlator $\phi_{q}^{\rm T}(t)$ have been derived in 
Ref.~\cite{Goetze91b}. 
The Zwanzig-Mori equation is given by
\begin{subequations}
\label{eq:GLE-T}
\begin{equation}
\phi_{q}^{\rm T}(t) + 
(\Omega_{q}^{\rm T})^{2} \int_{0}^{t} dt^{\prime} 
m_{q}^{\rm T}(t-t^{\prime}) \, \phi_{q}^{\rm T}(t^{\prime}) = 0,
\label{eq:GLE-T-a}
\end{equation}
and its Fourier-Laplace transform reads
\begin{equation}
\phi_{q}^{\rm T}(\omega) = - 
\frac{1}{\omega + (\Omega_{q}^{\rm T})^{2} \, m_{q}^{\rm T}(\omega)},
\label{eq:GLE-T-b}
\end{equation}
\end{subequations}
with $(\Omega_{q}^{\rm T})^{2} = q^{2} v^{2}$.
The relaxation kernel $m_{q}^{\rm T}(t)$ is a correlation
function of transversal fluctuating forces.
Within the MCT, the transversal 
fluctuating forces are approximated by 
their projection onto the subspace spanned by the density products,
and with the factorization approximation
the kernel is given by a mode-coupling functional 
${\cal F}_{q}^{\rm T}$ of the density correlators
\begin{subequations}
\label{eq:MCT-T}
\begin{equation}
m_{q}^{\rm T}(t) = {\cal F}_{q}^{\rm T}[\phi(t)], \quad
{\cal F}_{q}^{\rm T}[\tilde{f}] =
\sum_{k,p}
V_{q,kp}^{\rm T} \, \tilde{f}_{k} \tilde{f}_{p}.
\label{eq:MCT-T-a}
\end{equation}
Here $V_{q,kp}^{\rm T} = \rho S_{k} S_{p} \,
[ \, {\vec e}^{\,\, \rm T} \cdot 
({\vec k} c_{k} + {\vec p} c_{p}) \,]^{2} \, / \,
(2 q^{2})$
with ${\vec e}^{\,\, \rm T}$ denoting a unit vector 
orthogonal to ${\vec q}$. 
The $q \to 0$ limit of the functional is given by
\begin{equation}
{\cal F}_{0}^{\rm T}[\tilde{f}] = \sum_{k} 
V_{k}^{\rm T} \tilde{f}_{k}^{2},
\label{eq:MCT-T-b}
\end{equation}
\end{subequations}
with $V_{k}^{\rm T} = (\rho/60 \pi^{2})
k^{4} \, [c_{k}^{\prime} S_{k}]^{2}$.
These MCT equations of motion for $\phi_{q}^{\rm T}(t)$
can be solved with the knowledge of the static quantities 
and of the density correlators. 

In glass states, $m_{q}^{\rm T}(\omega)$ acquires a 
zero-frequency pole
since the transversal force fluctuations 
do not decay to zero for long times,
$m_{q}^{\rm T}(t \to \infty) = C_{q}^{\rm T} > 0$.
We therefore write in analogy to 
Eq.~(\ref{eq:GLE-hat-phi-c})
\begin{equation}
m_{q}^{\rm T}(t) = C_{q} +
C_{q}^{\rm T} \, \hat{m}_{q}^{\rm T}(t), \quad
C_{q}^{\rm T} = {\cal F}_{q}^{\rm T}[f],
\label{eq:MCT-T-glass}
\end{equation}
and introduce $\hat{m}_{q}^{\rm T}(t)$
for the description of the glass-state transversal current dynamics. 
The new kernel satisfies
$\lim_{t \to \infty} \hat{m}_{q}^{\rm T}(t) = 0$ and
$\lim_{\omega \to 0} \omega \hat{m}_{q}^{\rm T}(\omega) = 0$. 
Substituting the Fourier-Laplace transform of 
Eq.~(\ref{eq:MCT-T-glass}) into Eq.~(\ref{eq:GLE-T-b}) yields 
for the transversal current spectrum 
\begin{equation}
\phi_{q}^{{\rm T}''}(\omega) =
\frac{\omega^{2} (\hat{\Omega}_{q}^{\rm T})^{2}  
\hat{m}_{q}^{{\rm T}''}(\omega)}
{[\omega^{2} - (\hat{\Omega}_{q}^{\rm T})^{2} 
\hat{\Delta}_{q}^{\rm T}(\omega)]^{2} +
[\omega (\hat{\Omega}_{q}^{\rm T})^{2} 
\hat{m}_{q}^{{\rm T}''}(\omega)]^{2}},
\label{eq:GLE-T-glass-b}
\end{equation}
where
$\hat{\Omega}_{q}^{\rm T} = \Omega_{q}^{\rm T} \sqrt{C_{q}^{\rm T}}$
and $\hat{\Delta}_{q}^{\rm T} 
= 1 - \omega \, \hat{m}_{q}^{{\rm T}'}(\omega)$.

The reformulated equation of motion for the 
glass-state transversal current dynamics
can be closed by finding an expression for 
$\hat{m}_{q}^{\rm T}(t)$ as a new mode-coupling functional
$\hat{\cal F}_{q}^{\rm T}$. 
One finds from Eqs.~(\ref{eq:GLE-hat-phi-a}),
(\ref{eq:MCT-T-a}), and (\ref{eq:MCT-T-glass})
that 
$\hat{\cal F}_{q}^{\rm T} = 
\hat{\cal F}_{q}^{{\rm T} \, (1)} + 
\hat{\cal F}_{q}^{{\rm T} \, (2)}$
in analogy to Eq.~(\ref{eq:MCT-hat-phi-a}).
Correspondingly, there holds
$\hat{m}_{\rm T}(t) = 
\hat{m}_{\rm T}^{(1)}(t) + \hat{m}_{\rm T}^{(2)}(t)$
for $\hat{m}_{\rm T}(t) = \hat{m}^{\rm T}_{q=0}(t)$.
Explicit expressions for the new mode-coupling functionals
shall be omitted here for brevity:
they take the same form as in 
Eqs.~(\ref{eq:MCT-hat-phi}) and (\ref{eq:MCT-hat-phi-qnull}),
but with coupling coefficients replaced by the ones for
the transversal current correlators. 

The mathematical form of Eq.~(\ref{eq:GLE-T-glass-b})
for glass states
is identical to that for the longitudinal current 
given in Eq.~(\ref{eq:GLE-L-glass}). 
In particular, $\phi_{q}^{{\rm T}''}(\omega)$ for
glass states vanishes
proportional to $\omega^{2}$ for small frequencies
in contrast to the one in liquid states.
Thus, the transversal current correlators exhibit
a drastic difference between liquid and glass 
states~\cite{Goetze91b}: 
the long wave length correlators in liquids describe 
diffusive processes, 
while the ones in the glass states describe the 
propagation of transversal sound waves 
as expected for an isotropic elastic continuum.

\subsubsection{Velocity correlator}

Finally, we briefly summarize here for later discussion 
the MCT equations of motion for the
normalized velocity correlator
$\Psi(t) = \langle 
{\vec v}_{s}(t) \cdot {\vec v}_{s}(0) \rangle \, / \, (3v^{2})$
defined with the velocity ${\vec v}_{s}$
of a tagged particle (labeled $s$).
An exact equation for this quantity is given by
\begin{equation}
\partial_{t} \Psi(t) + 
v^{2} \int_{0}^{t} dt^{\prime} \,
m_{s}(t-t^{\prime}) \, \Psi(t^{\prime}) = 0,
\end{equation}
and the MCT expression for the kernel $m_{s}(t)$ reads
\begin{equation}
m_{s}(t) = \sum_{k} V_{k}^{s} \, \phi_{k}^{s}(t) \, \phi_{k}(t),
\end{equation}
with $V_{k}^{s} = (\rho / 6 \pi^{2}) 
k^{4} c_{k}^{2} S_{k}$~\cite{Goetze91b,Chong01b}.
Here $\phi_{q}^{s}(t) = 
\langle \rho_{\vec q}^{s}(t)^{*} \rho_{\vec q}^{s}(0) \rangle$
with $\rho_{\vec q}^{s} = \exp(i {\vec q} \cdot {\vec r}_{s})$
denotes tagged-particle density correlator.
The Zwanzig-Mori equation for $\phi_{q}^{s}(t)$ 
has the same form as Eq.~(\ref{eq:GLE-phi-a})
with $\phi_{q}$, $m_{q}$, and $\Omega_{q}^{2}$ replaced by
$\phi_{q}^{s}$, $m_{q}^{s}$, and $(\Omega_{q}^{s})^{2} = q^{2} v^{2}$,
respectively, and the MCT kernel is given by the functional
$m_{q}^{s}(t) = \sum_{k,p} V_{q,kp}^{s} \phi_{k}^{s}(t) \phi_{p}(t)$
with $V_{q,kp}^{s}$ determined by $S_{q}$~\cite{Bengtzelius84,Fuchs98}. 
For glass states, the MCT equations for $\Psi(t)$ can be
reformulated in terms of the hatted kernel
$\hat{m}_{s}(t)$ 
defined in analogy to Eq.~(\ref{eq:MCT-T-glass}) via
\begin{equation}
m_{s}(t) = C_{s} + C_{s} \hat{m}_{s}(t),
\,\,\,\,
C_{s} = \lim_{t \to \infty} m_{s}(t).
\end{equation}

\begin{figure}[htb]
\includegraphics[width=0.8\linewidth]{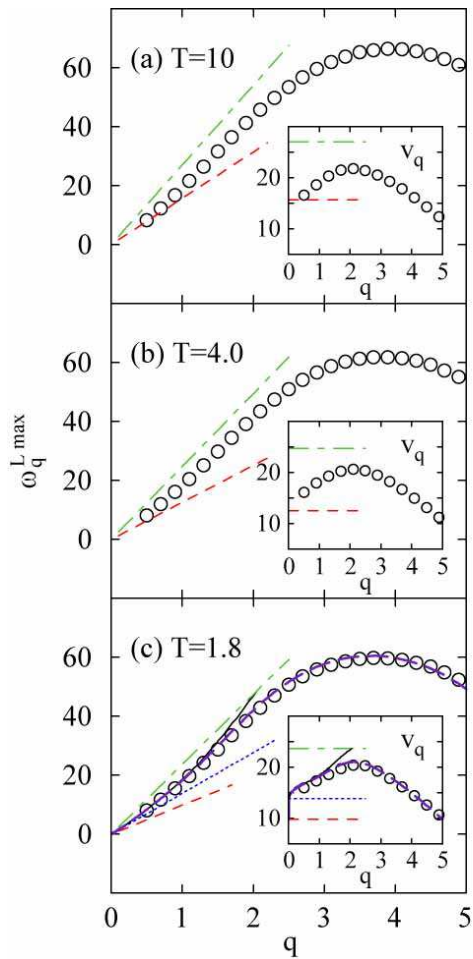}
\caption{
Peak positions $\omega_{q}^{\rm L \, max}$ (circles)
of the longitudinal current
spectra $\phi_{q}^{{\rm L} ''}(\omega)$ 
as a function of $q$ at $\rho = 1.093$
for temperatures indicated
in each panel referring to liquid states. 
Dashed lines denote the hydrodynamic dispersion law $q v_{0}$
with the isothermal sound velocity $v_{0}$.
Dash-dotted lines show the linear dispersion law
$q v_{\infty}$ with the infinite-frequency sound
velocity $v_{\infty}$. 
The insets exhibit apparent sound velocities
$v_{q} = \omega_{q}^{\rm L \, max} /q$ (circles) as a function of $q$
along with horizontal lines denoting 
$v_{0}$ (dashed lines) and
$v_{\infty}$ (dash-dotted lines).
Dotted lines in (c) 
refer to the linear dispersion law $q v_{\infty}^{\alpha}$ and
the velocity $v_{\infty}^{\alpha}$ discussed in the
text ({\em cf.} Sec.~\ref{sec:sound-4}).
Long-dashed lines through circles in (c)
are from the solution to Eq.~(\ref{eq:lowest-order-GHD}),
while solid lines are based on 
Eqs.~(\ref{eq:positive-dispersion}).}
\label{fig:dispersion-liquids}
\end{figure}

\section{Anomalous dispersion of sound-velocity}
\label{sec:sound}

\subsection{Dispersion relation}
\label{sec:sound-1}

Circles in Fig.~\ref{fig:dispersion-liquids} show 
the dispersion relation -- 
the peak positions $\omega_{q}^{\rm L \, max}$ 
of the longitudinal current
spectra $\phi_{q}^{{\rm L} ''}(\omega)$ 
as a function of wave number $q$ --
for three representative temperatures 
referring to liquid states.
(We notice that, in experimental and 
computer-simulation studies
for the dispersion relation cited in Sec.~\ref{sec:introduction},
it is the peak positions of the longitudinal current spectra that 
are usually reported.
The peak positions of the dynamic structure factors
\protect$\propto \phi_{q}''(\omega)$ and of the longitudinal current
spectra \protect$\phi_{q}^{{\rm L}''}(\omega)$ nearly coincide
as can be inferred, e.g., from Fig.~\protect\ref{fig:Sqw-CL-CT}.
In the present article, we are primarily interested in the
wave-number regime $q \lesssim 4$ 
where $\omega_{q}^{\rm L \, max}$ increases with $q$.
In this pseudo first Brillouin zone, the peaks
$\omega_{q}^{\rm L \, max}$ are associated with the
collective mode.)
The dashed line in each panel
denotes the hydrodynamic dispersion law
$q v_{0}$ with the isothermal sound velocity $v_{0}$~\cite{comment-v0},
whereas the dash-dotted line exhibits the linear
dispersion law $q v_{\infty}$ defined in terms of the
infinite-frequency sound velocity $v_{\infty}$ (see below).
The inset in each panel shows an apparent sound velocity
$v_{q} = \omega_{q}^{\rm L \, max}/q$ (circles)
as a function of $q$, 
along with horizontal lines denoting $v_{0}$ (dashed line) and
$v_{\infty}$ (dash-dotted line). 
The increase of the sound velocity $v_{q}$ from 
$v_{0}$ towards $v_{\infty}$ upon increase of the
wave number $q$, which is observable for $q \lesssim 2$ 
in all the insets of Fig.~\ref{fig:dispersion-liquids}, 
is called the anomalous or positive dispersion.

There are subtle variations in the anomalies depending
on the temperature.
At high temperature $T = 10$, 
the whole increase of the
sound velocity is observable in Fig.~\ref{fig:dispersion-liquids}(a)
starting from $v_{0}$ at small $q$ 
up to close to $v_{\infty}$ at $q \approx 2$. 
This feature is altered if the temperature is decreased. 
For $T = 4.0$ (Fig.~\ref{fig:dispersion-liquids}(b)),
the sound velocity $v_{q}$ for small wave numbers is
still located above the hydrodynamic value $v_{0}$. 
For $T = 1.8$ (Fig.~\ref{fig:dispersion-liquids}(c)),
this ``gap'' becomes larger, 
and the sound velocity $v_{q}$ in the
small-$q$ limit seems to approach another limit denoted by the
dotted line which will be defined below. 

\begin{figure}[tb]
\includegraphics[width=0.8\linewidth]{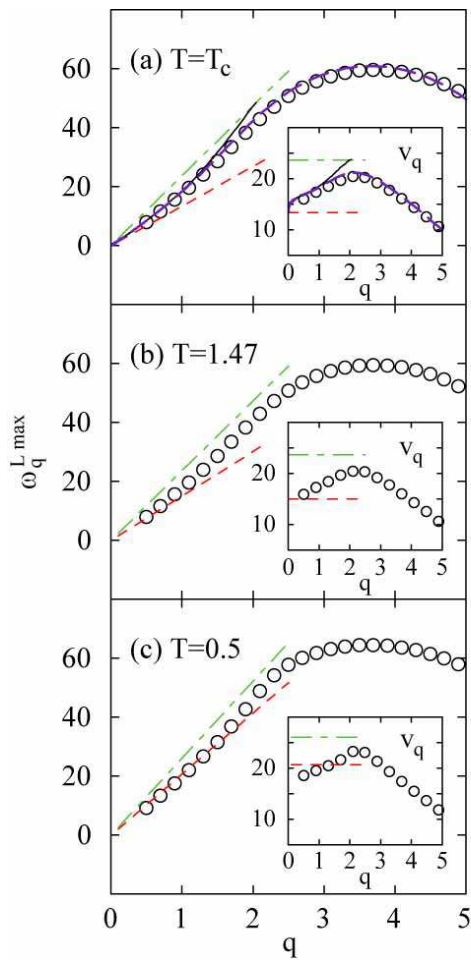}
\caption{
The same as in Fig.~\ref{fig:dispersion-liquids}, but here
temperatures refer to glass states, and 
the hydrodynamic dispersion law 
$q \hat{v}_{0}$ is determined by 
$\hat{v}_{0}$ defined in the text.
Long-dashed lines through circles in (a)
are from the solution to Eq.~(\ref{eq:lowest-order-GHD}),
while solid lines are based on
Eqs.~(\ref{eq:positive-dispersion}).}
\label{fig:dispersion-glasses}
\end{figure}

Corresponding results for representative glass states are
shown in Fig.~\ref{fig:dispersion-glasses}. 
But here, the hydrodynamic dispersion law  
$q \hat{v}_{0}$ is determined by 
$\hat{v}_{0}$ introduced in connection with
Eq.~(\ref{eq:GLE-hat-phi-b}) (see also below).
It is clear from Fig.~\ref{fig:dispersion-glasses}
that the anomalous sound-velocity dispersion
is present also in glass states.
Again, there are subtle variations in the 
appearance of the anomalies. 
For $T = 1.47$ (Fig.~\ref{fig:dispersion-glasses}(b)),
the sound velocity $v_{q}$ approaches
the hydrodynamic value $\hat{v}_{0}$ rather rapidly
for $q \to 0$. 
On the other hand, 
for $T = T_{\rm c}$ (Fig.~\ref{fig:dispersion-glasses}(a)),
the approach towards $\hat{v}_{0}$ is retarded. 
For the deep-in-glass state $T = 0.5$ 
(Fig.~\ref{fig:dispersion-glasses}(c)),
a new feature shows up, and 
``negative'' dispersion is observable for $q < 1$,
where the sound velocity $v_{q}$ is smaller than
the hydrodynamic value $\hat{v}_{0}$. 

The results shown in Figs.~\ref{fig:dispersion-liquids}(b),
\ref{fig:dispersion-liquids}(c), 
and \ref{fig:dispersion-glasses}(a) indicate that
the increase of the sound velocity from the hydrodynamic
value has already started at smaller wave-number regime which 
cannot be adequately resolved with the linear $q$ axis.
In addition, the presence of the negative dispersion 
shown in Fig.~\ref{fig:dispersion-glasses}(c)
implies that the anomalous sound-velocity dispersion 
cannot be accounted for solely in terms of 
relaxation processes since those processes would lead only to 
the positive dispersion as known, e.g., from the
viscoelastic model. 

In the following, a microscopic and unified understanding 
of the mentioned anomalies shall be attempted.
This will be done based on 
the generalized hydrodynamic description~\cite{Goetze00}.
In this description, the memory kernel $m_{q}(\omega)$ is 
approximated by its $q \to 0$ limit, while 
the full $\omega$ dependence of $m(\omega) = m_{q=0}(\omega)$
is retained.
In this manner, one gets from 
Eq.~(\ref{eq:GLE-L-b})~\cite{comment-GHD}
\begin{equation}
\phi_{q}^{{\rm L}''}(\omega) \approx
\frac{\omega^{2} \, \Omega_{q}^{2} \, m''(\omega)}
{[\omega^{2} - \Omega_{q}^{2} \, \Delta(\omega)]^{2} +
[\omega \, \Omega_{q}^{2} \, m''(\omega)]^{2}},
\label{eq:GLE-L-GHD}
\end{equation}
with $\Delta(\omega) = 1 - \omega m'(\omega)$. 
In lowest order, the resonance frequency 
$\omega_{q}^{\rm L \, max}$ can be obtained
as a solution to 
\begin{equation}
\omega_{q}^{\rm L \, max} = \Omega_{q} \, 
\sqrt{\Delta(\omega = \omega_{q}^{\rm L \, max})}.
\label{eq:lowest-order-GHD}
\end{equation}
The long-dashed lines through circles in 
Figs.~\ref{fig:dispersion-liquids}(c) and
\ref{fig:dispersion-glasses}(a) 
show the dispersion law and the sound velocity
based on the solution to this equation.
One understands that the lowest-order solution describes
the results from the full MCT solutions (circles) fairly well for the
whole $q$-range shown in the figure. 
Since we are interested in the small wave-number region
$q \lesssim 2$ where the anomalous sound-velocity dispersion
is present, a further simplification is possible:
one can replace $\Omega_{q}$ by its leading-order contribution,
$\Omega_{q} \approx q v_{0}$. 
This yields 
\begin{subequations}
\label{eq:positive-dispersion}
\begin{equation}
\omega_{q}^{\rm L \, max} = 
q v_{0} \, \sqrt{\Delta(\omega = \omega_{q}^{\rm L \, max})},
\label{eq:positive-dispersion-a}
\end{equation}
which can be rewritten for the 
the sound velocity $v_{q}$ as
\begin{equation}
v_{q} = v_{0} \sqrt{\Delta(\omega = q v_{q})}.
\label{eq:positive-dispersion-b}
\end{equation}
\end{subequations}
The solid lines in 
Figs.~\ref{fig:dispersion-liquids}(c) and
\ref{fig:dispersion-glasses}(a) exhibit
the dispersion law and the sound velocity 
based on the solutions to these equations. 
One understands that 
no serious error is introduced with the use of 
Eqs.~(\ref{eq:positive-dispersion})
as far as the regime $q \lesssim 2$ of interest
is concerned. 

Before proceeding, let us show that Eqs.~(\ref{eq:positive-dispersion})
reproduce known results~\cite{Hansen86,Balucani94}
in the limit of small and large resonance frequencies. 
The hydrodynamic sound is obtained by taking 
the $\omega \to 0$ limit of $\Delta(\omega)$ in
Eqs.~(\ref{eq:positive-dispersion}). 
One obtains the dispersion law 
$\omega_{q}^{\rm L \, max} = q v_{0} \sqrt{1 + m(t \to \infty)}$
since $\lim_{\omega \to 0} \omega m'(\omega) = 
- \lim_{t \to \infty} m(t)$. 
For liquids where the memory kernel relaxes to zero for long times,
$m(t \to \infty) = 0$, 
the dispersion law 
$\omega_{q}^{\rm L \, max} = q v_{0}$ is obtained with 
the isothermal sound velocity $v_{0}$~\cite{comment-v0}.
For glasses, on the other hand,
the kernel does not decay to zero for long times,
$m(t \to \infty) = C$, leading to
the dispersion law 
$\omega_{q}^{\rm L \, max} = q \hat{v}_{0}$ 
with $\hat{v}_{0} = v_{0} \sqrt{1 + C}$.
(This agrees with 
$\hat{v}_{0} = v_{0} / \sqrt{1 - f_{0}}$ given below
Eq.~(\ref{eq:GLE-hat-phi-b}) because of 
Eq.~(\ref{eq:DW}).)
These results have been included
with dashed lines 
in Figs.~\ref{fig:dispersion-liquids} and \ref{fig:dispersion-glasses}.
In the high-frequency limit, one obtains, 
using $\lim_{\omega \to \infty} \Delta(\omega) = 1 + \lim_{t \to 0}m(t)$,
the dispersion law
$\omega_{q}^{\rm L \, max} = q v_{\infty}$
with the infinite-frequency sound velocity
$v_{\infty} = v_{0} \sqrt{1 + m(t=0)}$.
This holds for both liquid and glass states,
and has been included with dash-dotted lines in 
Figs.~\ref{fig:dispersion-liquids} and \ref{fig:dispersion-glasses}.

Thus, on the basis of Eq.~(\ref{eq:positive-dispersion-b}), 
it is clear that the anomalous sound-velocity
dispersion from the low-frequency hydrodynamic value
($v_{0}$ for liquids and $\hat{v}_{0}$ for glasses) towards
the infinite-frequency one $v_{\infty}$ upon 
increase of the wave number $q$
is controlled by the detailed frequency dependence of $\Delta(\omega)$.
This in turn is determined by the relaxation of the memory
kernel $m(t)$. 
Within the MCT, $m(t)$ is given in terms of the mode-coupling
functional as $m(t) = {\cal F}_{0}[\phi(t)]$ 
[{\em cf.} Eq.~(\ref{eq:MCT-phi-c})].
In the following two subsections, the details of the
relaxation of $m(t)$ and of the frequency dependence of
$\Delta(\omega)$ will be investigated. 
These results will be employed 
in Sec.~\ref{sec:sound-4} for a systematic
study of the anomalies. 

\subsection{Structural-relaxation processes}
\label{sec:sound-2}

\begin{figure}[tb]
\includegraphics[width=0.8\linewidth]{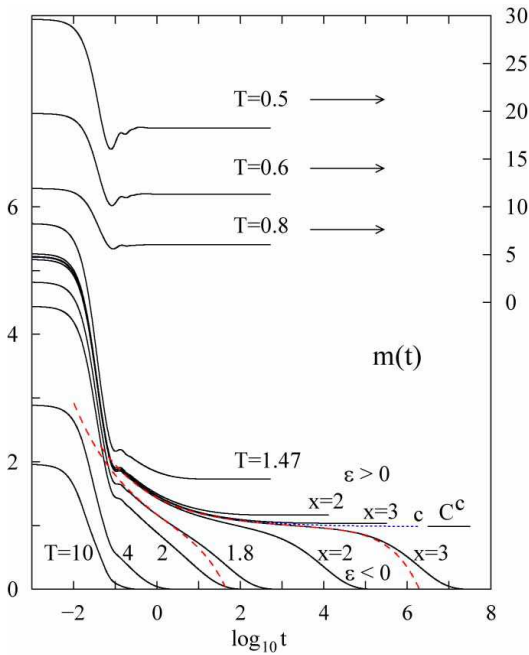}
\caption{
The memory kernel at $q=0$, $m(t) = m_{q=0}(t)$,
as a function of $\log_{10}t$  
at $\rho = 1.093$ for temperatures indicated in the figure.
Curves with labels $x=2$ and 3 are calculated for
distance parameters 
$\epsilon = (T_{\rm c} - T)/T_{\rm c} = \mp 10^{-x}$
for liquids ($\epsilon < 0$) and 
for glasses ($\epsilon > 0$). 
Dotted line with label ${\rm c}$ denotes the kernel at $T_{\rm c}$.
Horizontal line marks the plateau height $C^{\rm c}$. 
Dashed lines show the MCT asymptotic-law results
based on Eq.~(\ref{eq:first-m})
for $T = 1.8$ and $x=3$ ($\epsilon < 0$). 
The kernels for $T = 0.5$, 0.6, and 0.8 refer to the right vertical axis.}
\label{fig:zm-qnull}
\end{figure}

Figure~\ref{fig:zm-qnull} exhibits the evolution of the dynamics
of $m(t)$ upon decrease of the temperature $T$.
The curve for $T = 10$, which is quite higher than 
$T_{\rm c} \approx 1.637$, decays rapidly 
and nearly exponentially to zero.
By lowering the temperature, 
the dynamics becomes slower and stretched 
due to the development of the cage effect. 
For temperatures close to but above $T_{\rm c}$,
the kernel $m(t)$ exhibits a two-step relaxation:
the dynamics towards the plateau $C^{\rm c}$, 
followed by the final relaxation from the plateau to zero.
The dynamics that occurs near the plateau is referred to
as the $\beta$ process, whereas the final relaxation is 
called the $\alpha$ process ({\em cf.} Sec.~\ref{sec:theory-3}).
The $\beta$ process for small $|\, T - T_{\rm c} \,|$ can be
well described by the MCT asymptotic formula~(\ref{eq:beta-correlator-a})
specialized to $m(t)$
\begin{equation}
m(t) = C^{\rm c} + D G(t),
\label{eq:first-m}
\end{equation}
as exemplified with the dashed lines in Fig.~\ref{fig:zm-qnull}.
($C^{\rm c} = 0.982$ and $D = 1.925$ for the model under study.)
At the critical point $T_{\rm c}$, the kernel approaches the plateau
in a stretched manner, 
as shown by the dotted curve with label c
in Fig.~\ref{fig:zm-qnull}.
This process, called the critical decay, is described by 
a power law [{\em cf.} Eq.~(\ref{eq:critical-X})]
\begin{equation}
m(t) = C^{\rm c} + D (t_{0}/t)^{a}. 
\label{eq:critical-m}
\end{equation}
Decreasing $T$ below the critical temperature $T_{\rm c}$,
the values for the long-time limits $C$ increase.
These limits are approached exponentially fast for
$T \ne T_{\rm c}$~\cite{Goetze95b}.
If $T$ is far below $T_{\rm c}$, 
one observes enhanced oscillatory features 
for times near $\log_{10} t \approx -1.2$ 
as can be seen from the curves for $T = 0.8$, 0.6, and 0.5
in Fig.~\ref{fig:zm-qnull}.
Such vibrational dynamics reflects 
the anomalous-oscillation peak (AOP)
discussed in Ref.~\cite{Goetze00}.
It is caused by strong interaction between density 
fluctuations at microscopic length scales and the 
arrested glass structure.
Features of the AOP will be discussed in the next subsection.

\begin{figure}[tb]
\includegraphics[width=0.8\linewidth]{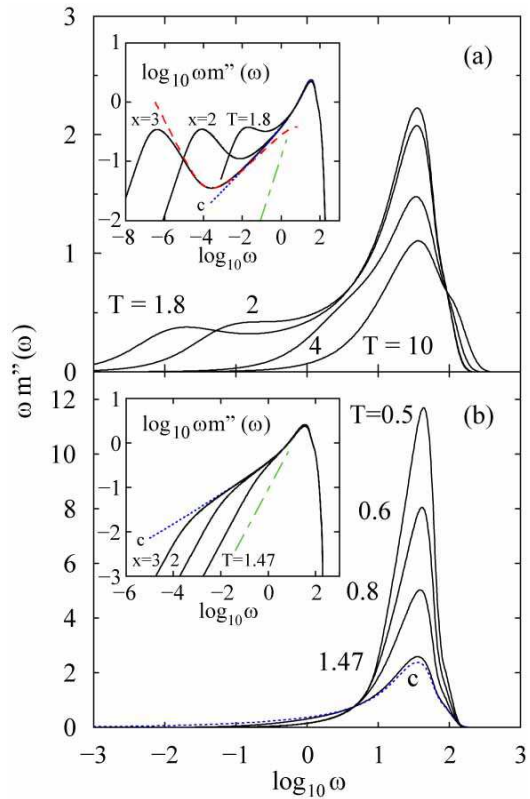}
\caption{
Susceptibility spectra $\omega m''(\omega)$ 
for the memory kernels $m(t)$ shown in Fig.~\ref{fig:zm-qnull}
for liquid states (a) and for glass states (b).
The insets exhibit the double logarithmic representation
of the spectra for $T$ close to $T_{\rm c}$.
The meaning of the labels $x=2$ and $x=3$ is the same
as in Fig.~\ref{fig:zm-qnull}. 
Dotted line with label ${\rm c}$ denotes the 
susceptibility spectrum at $T_{\rm c}$.
Dashed line in the upper inset shows the MCT asymptotic-law result
based on Eq.~(\ref{eq:first-susceptibility})
for $x=3$ ($\epsilon < 0$). 
Dash-dotted lines in the insets exhibit the hydrodynamic
linear law $ \sim \omega$.}
\label{fig:susceptibility-zm-qnull}
\end{figure}

Figure~\ref{fig:susceptibility-zm-qnull}
shows the susceptibility spectra $\omega m''(\omega)$ of the kernels
shown in Fig.~\ref{fig:zm-qnull}.
The spectrum at $T = 10$ exhibits only one peak located
at high frequencies, to be called a microscopic peak.
By lowering $T$, an additional peak emerges at low frequencies
reflecting the evolution of the 
structural-relaxation processes.
The lower frequency peak, whose position decreases 
strongly by lowering $T$, is due to the $\alpha$ process,
and is called the $\alpha$ peak.
It is separated from the microscopic peak by a minimum,
called the $\beta$ minimum.
The appearance of the minimum is caused by the 
crossover from a power law $\sim t^{-a}$ to another one
$\sim - t^{b}$ which occurs in the $\beta$ regime 
({\em cf.} Sec.~\ref{sec:theory-3}).
For small $|\, T - T_{\rm c} \,|$,
the $\beta$ minimum is well 
described by the formula which follows from 
Eq.~(\ref{eq:first-m}), 
\begin{equation}
\omega m''(\omega) = D \omega G''(\omega),
\label{eq:first-susceptibility}
\end{equation} 
as exhibited by the dashed line
in the inset of Fig.~\ref{fig:susceptibility-zm-qnull}(a).
At $T = T_{\rm c}$, the critical decay (\ref{eq:critical-m})
leads to a sublinear susceptibility variation 
\begin{equation}
\omega m''(\omega) = D \sin(\pi a / 2) \Gamma(1-a) 
(\omega t_{0})^{a},
\label{eq:critical-susceptibility}
\end{equation}
as shown by the dotted line with label c.
Here, $\Gamma(x)$ denotes the Gamma function. 
Below $T_{\rm c}$, the final exponential decay 
towards $C$~\cite{Goetze95b} 
implies the hydrodynamic law in the small-$\omega$ limit, 
$\omega m''(\omega \to 0) \propto \omega$.
But, near $T_{\rm c}$, there is a frequency interval 
where the spectrum is described by the 
sublinear critical spectrum (\ref{eq:critical-susceptibility}). 
This produces a ``knee'' at some frequency where
the crossover from the linear low-frequency spectrum
$\omega m''(\omega) \propto \omega$
to the sublinear critical one
$\omega m''(\omega) \propto \omega^{a}$ occurs,
as can be inferred from the inset of 
Fig.~\ref{fig:susceptibility-zm-qnull}(b). 
If $T$ is far below $T_{\rm c}$, the knee disappears,
and the spectrum is dominated by a single peak
reflecting the AOP (see the next subsection).
This holds for $T = 0.8$, 0.6, and 0.5.
It is interesting to notice that there is only a weak temperature
dependence in the position $\log_{10} \omega \approx 1.5$
of the high-frequency peak
from high-$T$ liquids down to deep-in-glass states. 
This feature will also be studied in the next subsection.

\begin{figure}[tb]
\includegraphics[width=0.8\linewidth]{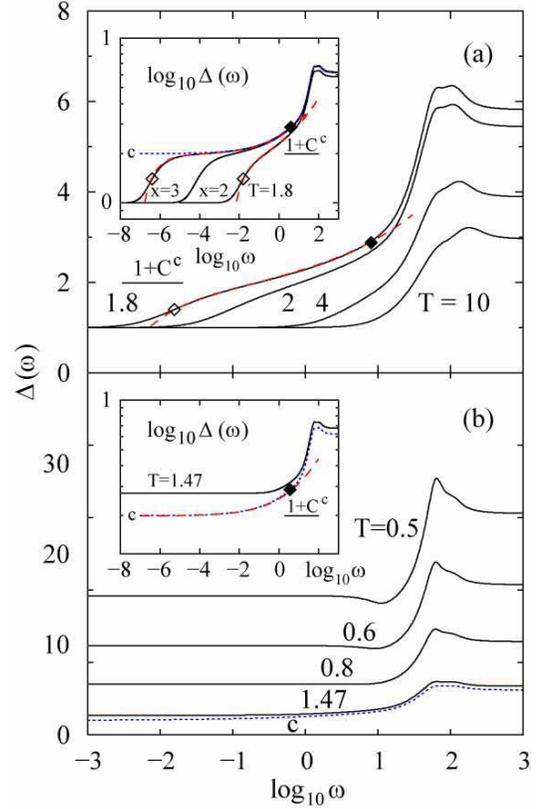}
\caption{
Reactive part of the normalized longitudinal moduli
$\Delta(\omega) = 1 - \omega m'(\omega)$ 
for the memory kernels $m(t)$ shown in Fig.~\ref{fig:zm-qnull}
for liquid states (a) and for glass states (b).
The insets exhibit the double logarithmic representation
of $\Delta(\omega)$ for $T$ close to $T_{\rm c}$.
The meaning of the labels $x=2$ and $x=3$ is the same
as in Fig.~\ref{fig:zm-qnull}. 
Dotted line with label ${\rm c}$ denotes $\Delta(\omega)$
at $T_{\rm c}$.
Horizontal lines mark the plateau height $1 + C^{\rm c}$
(see text).
Dashed lines in the upper inset show the MCT asymptotic-law results
based on Eq.~(\ref{eq:Delta-beta}) 
for $T = 1.8$ and $x=3$ ($\epsilon < 0$), 
while dashed line in the lower inset exhibits the curve
based on Eq.~(\ref{eq:critical-Delta}).
The open ($\omega_{\beta}^{-}$) 
and filled ($\omega_{\beta}^{+}$) diamonds 
refer to the frequencies $\omega_{\beta}^{\pm}$ defined in the text.}
\label{fig:Delta-zm-qnull}
\end{figure}

Figure~\ref{fig:Delta-zm-qnull} shows the reactive part
of the normalized longitudinal
moduli $\Delta(\omega) = 1 - \omega m'(\omega)$.
Since $m'(\omega)$ and $m''(\omega)$ are related via the 
Kramers-Kronig relation~\cite{Hansen86},
features for $\Delta(\omega)$ can be understood 
from those for the susceptibility spectra just mentioned.
In particular, a peak in $\omega m''(\omega)$ implies
a strong increase of $\Delta(\omega)$ near that peak frequency.
This way, one understands an almost single-step increase 
of $\Delta(\omega)$ at $\log_{10} \omega \approx 1.5$
for $T \gg T_{\rm c}$ and $T \ll T_{\rm c}$,
reflecting the presence of only the 
high-frequency microscopic peak in $\omega m''(\omega)$
for those temperatures 
({\em cf}. Fig.~\ref{fig:susceptibility-zm-qnull}).
For $T \gtrsim T_{\rm c}$, on the other hand,  
the presence of two peaks in $\omega m''(\omega)$
explains the two-step increase of 
$\Delta(\omega)$ shown in Fig.~\ref{fig:Delta-zm-qnull}(a). 
Because of the strong $T$ dependence of the $\alpha$-peak
frequency, the position where the low-$\omega$ variation of
$\Delta(\omega)$ occurs depends strongly on the
temperature. 
In addition, the sublinear susceptibility
variation $\omega m''(\omega) \propto \omega^{a}$,
present both in liquid and glass states near $T_{\rm c}$,
also leads to the enhancement of $\Delta(\omega)$.
Indeed, it follows from Eq.~(\ref{eq:critical-m}) that
\begin{equation}
\Delta(\omega) = 1 + C^{\rm c} + 
D \cos(\pi a / 2) \Gamma(1-a) 
(\omega t_{0})^{a},
\label{eq:critical-Delta}
\end{equation}
holds near the $\beta$ minimum for liquids or 
near the knee position for glasses. 
The curve based on this formula (dashed line) is 
compared with the memory kernel at the critical point
(dotted line) in the inset of Fig.~\ref{fig:Delta-zm-qnull}(b).

For later convenience, let us define the 
$\alpha$ and $\beta$ regimes for $\Delta(\omega)$. 
We start from liquid states. 
Since the memory kernel $m(t)$ 
decays from the plateau $C^{\rm c}$ to zero
in the $\alpha$ regime, corresponding
$\alpha$ regime for $\Delta(\omega)$ shall be defined 
as the frequency interval $0 < \omega < \omega_{\alpha}^{+}$
in which $\omega_{\alpha}^{+}$ satisfies 
$\Delta(\omega_{\alpha}^{+}) = 1 + C^{\rm c}$.
Thus, the $\alpha$ regime for $\Delta(\omega)$
is below the horizontal bar marking 
$1 + C^{\rm c}$ in Fig.~\ref{fig:Delta-zm-qnull}(a).
In the $\beta$ regime, the kernel $m(t)$ is well described 
by the MCT asymptotic formula~(\ref{eq:first-m})
({\em cf.} Fig.~\ref{fig:zm-qnull}). 
The corresponding $\beta$ regime shall be defined
as the one where $\Delta(\omega)$ 
can be well described by the asymptotic formula
\begin{equation}
\Delta(\omega) = 1 + C^{\rm c} -
D \omega G'(\omega),
\label{eq:Delta-beta}
\end{equation}
which follows from Eq.~(\ref{eq:first-m}).
The dashed lines in Fig.~\ref{fig:Delta-zm-qnull}(a)
show curves based on this formula. 
Let us introduce $\omega_{\beta}^{\pm}$ 
as the frequencies where $\Delta(\omega)$ (solid lines)
differ from the MCT asymptotic formula~(\ref{eq:Delta-beta}) 
(dashed lines) by 2\%.
These frequencies are marked by open ($\omega_{\beta}^{-}$)
and filled ($\omega_{\beta}^{+}$) diamonds in 
Fig.~\ref{fig:Delta-zm-qnull}(a), and 
the frequency interval 
$\omega_{\beta}^{-} < \omega < \omega_{\beta}^{+}$
shall be considered 
as the $\beta$ regime for $\Delta(\omega)$.
As mentioned in Sec.~\ref{sec:theory-3}, 
there is an overlap between the $\alpha$ and $\beta$ regimes.
For glass states, 
the $\beta$ regime for $\Delta(\omega)$
shall be defined as the frequency interval
$\omega < \omega_{\beta}^{+}$ where 
$\omega_{\beta}^{+}$ marks the point
at which $\Delta(\omega)$ differs 
from the MCT asymptotic formula~(\ref{eq:critical-Delta})
by 2\%.
The frequency $\omega_{\beta}^{+}$ at the critical point
is marked by the filled diamond in the inset of 
Fig.~\ref{fig:Delta-zm-qnull}(b).

\subsection{Vibrational excitations}
\label{sec:sound-3}

In contrast to the structural-relaxation processes,
there is only a weak temperature dependence in the
time scale for the short-time dynamics in $m(t)$
occurring at $\log_{10} t \lesssim -1$ 
({\em cf}. Fig.~\ref{fig:zm-qnull}).
Correspondingly, the high-frequency
peaks in the susceptibility 
spectra $\omega m''(\omega)$ are located nearly
at the same frequencies $\log_{10} \omega \approx 1.5$ 
from high-$T$ liquid down to 
deep-in-glass state 
({\em cf}. Fig.~\ref{fig:susceptibility-zm-qnull}),
leading to a common increase of $\Delta(\omega)$
around this frequency regime 
({\em cf}. Fig.~\ref{fig:Delta-zm-qnull}).
In this section, nature of such high-frequency
microscopic process shall be discussed. 

\begin{figure}[tb]
\includegraphics[width=0.8\linewidth]{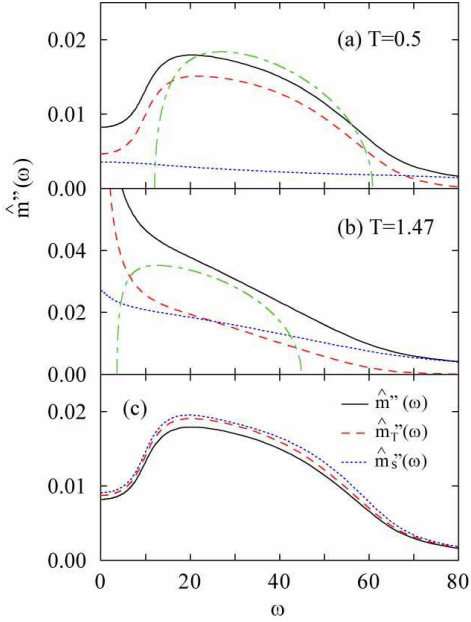}
\caption{
(a) Solid line denotes the memory-kernel
spectrum $\hat{m}''(\omega)$
for $\rho = 1.093$ and $T = 0.5$,
and dashed and dotted lines respectively refer to its
one-mode and two-mode 
contributions, $\hat{m}^{(1)''}(\omega)$ and $\hat{m}^{(2)''}(\omega)$.
Dash-dotted line shows the SGA spectrum 
$\hat{m}_{\rm SGA}''(\omega)$ based on 
Eqs.~(\ref{eq:harmonic-app}).
(b) The same as in (a), but for $T = 1.47$.
(c) Comparison of the memory-kernel spectrum $\hat{m}(\omega)$
(solid line) with $\hat{m}_{\rm T}''(\omega)$
(dashed line) for the transversal current correlator
and $\hat{m}_{s}''(\omega)$ (dotted line)
for the velocity correlator for $\rho = 1.093$ and $T = 0.5$.}
\label{fig:hat-memories-T050}
\end{figure}

We start from temperatures far below $T_{\rm c}$,
$T \ll T_{\rm c}$,
called stiff-glass states,
which are characterized by the Debye-Waller factors close
to unity as exemplified by the curve for $T = 0.5$ shown
in Fig.~\ref{fig:fq-Sq}.
One understands from Eqs.~(\ref{eq:MCT-hat-phi-b}) 
(\ref{eq:MCT-hat-phi-c}),
(\ref{eq:MCT-hat-phi-e}), and
(\ref{eq:MCT-hat-phi-f}) that the
renormalized coupling coefficients become very small
for the stiff-glass states, and they decrease towards
zero in the limit $\eta = 1 - f_{q} \to 0$.
Furthermore, the two-mode contributions to the kernels 
get suppressed relative
to the one-mode contributions, 
in particular 
$\hat{m}^{(2)}(t) / \hat{m}^{(1)}(t) = O(\eta)$.
Figure~\ref{fig:hat-memories-T050}(a) shows 
the hatted spectrum $\hat{m}''(\omega)$ for $T = 0.5$ (solid line)
along with its one-mode $\hat{m}^{(1)''}(\omega)$
(dashed line) and two-mode $\hat{m}^{(2)''}(\omega)$
(dotted line) contributions ({\em cf.} Eq.~(\ref{eq:MCT-hat-phi-d})). 
It is seen that the one-mode contribution
$\hat{m}^{(1)''}(\omega)$ indeed provides
the dominant contribution to 
the total memory-kernel spectrum
$\hat{m}''(\omega)$. 
The role played by the two contributions
$\hat{m}^{(1)}$ and $\hat{m}^{(2)}$
is utterly different.
The former describes the AOP,
while the latter provides a background spectrum.

Further properties of the AOP have been discussed 
in Ref.~\cite{Goetze00} based on a 
stiff-glass approximation (SGA),
which is obtained by dropping the two-mode contributions
to the memory kernels.
(See also Ref.~\cite{Thomas02} on the SGA.)
In particular, it has been found that harmonic oscillations
of the particles within their
cages are a good description of the relevant modes,
and the density correlator as well
as the memory kernel in the stiff-glass states
can be described as a superposition of
independent harmonic oscillator spectra
where the distribution of oscillator frequencies is
caused by the distribution 
of sizes and shapes of the cages. 
According to the cited MCT result for the AOP,
one obtains 
$\hat{m}(\omega) \approx 
\hat{m}_{\rm SGA}(\omega)$ with 
\begin{subequations}
\label{eq:harmonic-app}
\begin{equation}
\omega \hat{m}_{\rm SGA}(\omega) = 
w_{1} \, [ \, \tilde{\chi}(\omega) - 1 \, ].
\label{eq:harmonic-app-a}
\end{equation}
Here $\tilde{\chi}(\omega)$ is given by a superposition
of undamped harmonic-oscillator spectra
\begin{eqnarray}
& &
\tilde{\chi}(\omega) = \int_{\omega_{-}^{2}}^{\omega_{+}^{2}} 
d\xi \, \tilde{\rho}(\xi) \, \chi_{\xi}(\omega),
\\
& &
\tilde{\rho}(\xi) =
\sqrt{(\omega_{+}^{2} - \xi) \, (\xi - \omega_{-}^{2})} \, / \,
(2 \pi w_{1}),
\\
& &
\chi_{\xi}(\omega) = - \tilde{\Omega}^{2} \, / \,
[ \, \omega^{2} - \tilde{\Omega}^{2} \xi + i \omega \nu \, ].
\label{eq:harmonic-app-d}
\end{eqnarray}
\end{subequations}
The weight distribution $\tilde{\rho}(\xi)$ for the
oscillators with frequency $\sqrt{\xi} \tilde{\Omega}$
extends from the low-frequency threshold
$\Omega_{-} = \omega_{-} \tilde{\Omega}$ to
the high-frequency one $\Omega_{+} = \omega_{+} \tilde{\Omega}$,
where $\omega_{\pm} = 1 \pm \sqrt{w_{1}}$ with an
integrated coupling coefficient
$w_{1} = \sum_{k} \hat{V}_{k}^{(1)}$
({\em cf.} Eq.~(\ref{eq:MCT-hat-phi-e})).
$\tilde{\Omega}$ denotes some average value of 
$\hat{\Omega}_{q}$ defined in Eq.~(\ref{eq:GLE-hat-phi-b})
over the $q$ range $q \gtrsim q_{\rm D}$~\cite{Goetze00}.
Here, $q_{\rm D} = (6 \pi^{2} \rho)^{1/3}$ denotes the 
Debye wave number, and $q_{\rm D} = 4.02$ for the
density $\rho = 1.093$. 
Since the $q$ range around the first peak position $q_{\rm max}$
of the static structure factor $S_{q}$ is known to constitute
the dominant contribution to the kernel~\cite{Bengtzelius84},
$\tilde{\Omega}$ in the present study is taken as an 
average of the first maximum and first minimum of 
$\hat{\Omega}_{q}$, which occurs at $q \approx q_{\rm D}$
and $q_{\rm max}$, respectively 
({\em cf.} Fig.~\ref{fig:hat-Omega}).
A friction term $\nu$ is necessary 
when one wants to incorporate the
neglected two-mode contributions in a perturbative 
manner~\cite{Goetze00},
but one may read the formulas with $\nu=0$ unless
stated otherwise. 
Figure~\ref{fig:hat-memories-T050}(a) demonstrates that
Eqs.~(\ref{eq:harmonic-app}) with $\nu = 0$ 
(dash-dotted line) describe qualitative features 
of the AOP of $\hat{m}''(\omega)$ for $T = 0.50$.

By increasing $T$, one observes enhanced intensity in 
$\hat{m}''(\omega)$ 
at low frequencies as shown for $T = 1.47$ 
in Fig.~\ref{fig:hat-memories-T050}(b). 
This is a precursor phenomenon of the glass melting
at the transition point.
Near $T_{\rm c}$, such enhancement for decreasing
frequencies can be described by a 
critical power law $\sim \omega^{a-1}$
({\em cf.} Eq.~(\ref{eq:critical-susceptibility})).
By substituting the critical decay of the density
correlators $\hat{\phi}_{q}(t) \sim t^{-a}$ 
into Eq.~(\ref{eq:MCT-hat-phi-qnull-b}), 
one understands that it is the linear term
$\hat{m}^{(1)}(t)$ that provides the critical decay of
$\hat{m}(t)$, whereas the quadratic term
$\hat{m}^{(2)}(t)$ serves only as a correction $O(t^{-2a})$.
This explains why in Fig.~\ref{fig:hat-memories-T050}(b)
the strong increase of the intensity
is dominated by $\hat{m}^{(1)''}(\omega)$ (dashed line).
Thus, the modes building the AOP are buried
under the tail of the central peak, and they show up
only as a shoulder. 
Nevertheless, it is remarkable that 
the SGA spectrum $\hat{m}_{\rm SGA}''(\omega)$
(dash-dotted line) 
succeeds to extract the buried AOP portion 
as demonstrated in Fig.~\ref{fig:hat-memories-T050}(b).
The shift of the AOP to lower frequencies upon increase
of the temperature reflects the widened size of the cages
so that particles ``bounce'' in their cages
with smaller average frequencies.
Indeed, the particles are localized in such tight
cages at $T = 0.5$ that the square root 
$\delta r = \sqrt{ \langle \delta r^{2} (t \to \infty) \rangle }$
of the long-time limit of the mean-squared 
displacement~\cite{Fuchs98} 
is only 5.0\% of the particle's LJ diameter, whereas 
it is increased to $\delta r = 0.139$ at $T = 1.47$. 
The shift is accompanied by an increase of the
inelastic spectrum including the AOP, 
reflecting the decrease of the elastic contribution
$\pi C \delta(\omega)$. 

\begin{figure}[tb]
\includegraphics[width=0.8\linewidth]{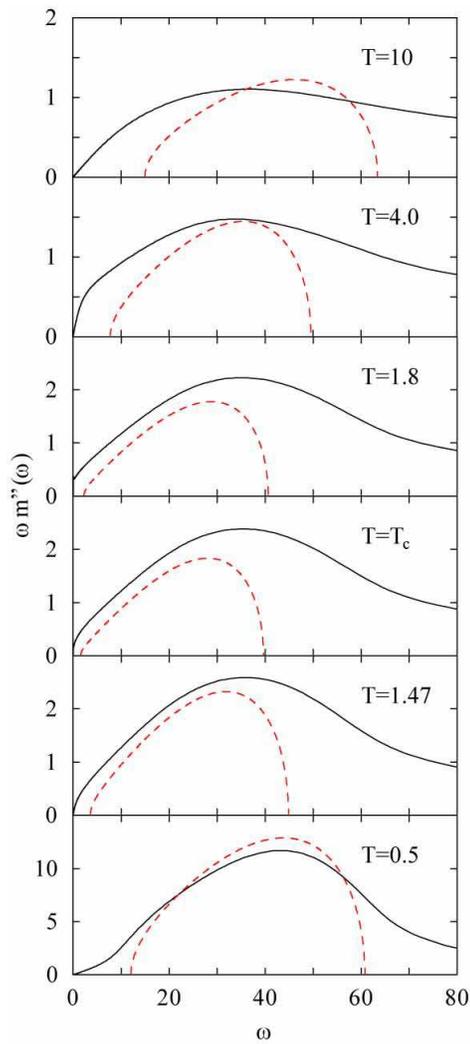}
\caption{
Susceptibility spectra 
$\omega m''(\omega)$ (solid lines) at indicated temperatures 
taken from 
Fig.~\ref{fig:susceptibility-zm-qnull},
but plotted here on the linear-$\omega$ axis. 
Dashed lines denote the SGA susceptibility spectra 
$\omega m_{\rm SGA}''(\omega)$ based on 
Eq.~(\ref{eq:susceptibility-harmonic-app}).}
\label{fig:susceptibility-memory-harmonic-app}
\end{figure}

The analysis presented in Fig.~\ref{fig:hat-memories-T050}(b)
also implies a possibility to extract AOP portion 
in the spectrum on the basis of Eqs.~(\ref{eq:harmonic-app}).
So, let us examine to what extent the 
high-frequency microscopic peaks in the susceptibility
spectra $\omega m''(\omega)$ at higher temperatures
have such ``harmonic'' character.
This is done in 
Fig.~\ref{fig:susceptibility-memory-harmonic-app} 
where representative susceptibility spectra taken from 
Fig.~\ref{fig:susceptibility-zm-qnull}
are compared with the SGA susceptibility spectra
\begin{equation}
\omega m_{\rm SGA}''(\omega) = (1 + C) \, w_{1} \, 
\tilde{\chi}''(\omega),
\label{eq:susceptibility-harmonic-app}
\end{equation}
which follows from 
Eqs.~(\ref{eq:GLE-hat-phi-c}) and (\ref{eq:harmonic-app-a}).
To focus on the high-frequency regime, the comparison is done
in Fig.~\ref{fig:susceptibility-memory-harmonic-app} 
with the linear-$\omega$ axis.
In evaluating the SGA susceptibility spectra 
for liquids,
nonergodicity parameters $f_{q}$ and $C$
in the equations and quantities involved are 
replaced by the ones at the critical point.

As discussed in connection with 
Figs.~\ref{fig:hat-memories-T050}(a) and
\ref{fig:hat-memories-T050}(b),
the SGA spectra describe the AOP portion of the memory-kernel
spectra for $T = 0.5$ and $T = 1.47$ fairly well, 
and this is reflected in the susceptibility spectra 
shown in the two bottom panels of 
Fig.~\ref{fig:susceptibility-memory-harmonic-app}.
Somewhat large deviation seen at $\omega \gtrsim 40$
for $T = 1.47$ is due to the neglected two-mode contribution,
which is more enhanced than the deviation in 
Fig.~\ref{fig:hat-memories-T050}(b) due to the presence
of the factor $\omega$ in $\omega m''(\omega)$. 
On the other hand, because of this factor, 
the deviation at small frequencies 
seen in Fig.~\ref{fig:hat-memories-T050}(b)
is suppressed for $\omega m''(\omega)$. 

One understands from the other panels of 
Fig.~\ref{fig:susceptibility-memory-harmonic-app} that
the harmonic approximation
based on Eq.~(\ref{eq:susceptibility-harmonic-app})
reasonably accounts for overall features 
-- peak position and strength -- 
of the microscopic peaks in $\omega m''(\omega)$ 
not only near $T_{\rm c}$ but even 
at high temperatures such as $T = 10$.
The large deviations for low frequencies at $T = 4.0$ and 10
simply reflect the fact that at such high $T$ 
the structural-relaxation contributions also 
enter into the high-frequency regime
({\em cf.} Fig.~\ref{fig:susceptibility-zm-qnull}).
Thus, considerable portion of the short-time or high-frequency
microscopic process, 
from high-$T$ liquids down to deep-in-glass states,
can be described as a superposition of harmonic 
vibrational dynamics. 
This explains the weak $T$ dependence of the microscopic process
shown in Figs.~\ref{fig:zm-qnull}-\ref{fig:Delta-zm-qnull}.

\begin{figure}[tb]
\includegraphics[width=0.8\linewidth]{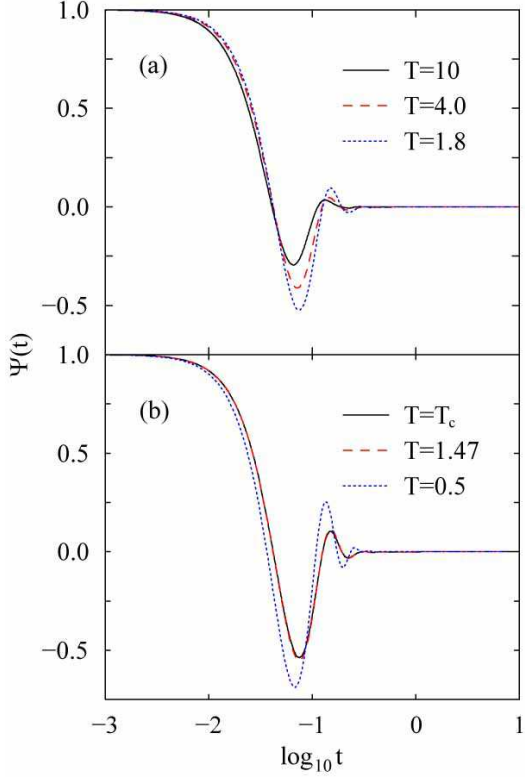}
\caption{
Normalized velocity correlator $\Psi(t)$ at $\rho = 1.093$
for temperatures referring to 
liquid states (a) and glass states (b).}
\label{fig:VACF}
\end{figure}

The persistence of the vibrational dynamics
inside the cage  
in the liquid state is not surprising.
To see this, we show in Fig.~\ref{fig:VACF}
the velocity correlator $\Psi(t)$ for
representative liquid and glass states. 
It is well known that the cage effect in 
dense liquids manifests itself
by oscillatory variations with a decay of $\Psi(t)$
to negative values~\cite{Hansen86},
and Fig.~\ref{fig:VACF}(a) demonstrates this phenomenon.
That such oscillatory dynamics exhibited by $\Psi(t)$ 
reflects the AOP can be understood from the observations
(i) in the stiff-glass state $T = 0.5$
the spectrum $\hat{m}_{s}''(\omega)$ 
of the relaxation kernel for $\Psi(t)$
exhibits an AOP whose spectral features -- threshold frequencies
and maximum position and height -- nearly coincide with 
those for the AOP of $\hat{m}''(\omega)$
as demonstrated in Fig.~\ref{fig:hat-memories-T050}(c),
and (ii) at higher $T$ including liquid states
there is a correlation between the peak position
in the SGA susceptibility spectra $\omega m_{\rm SGA}''(\omega)$
shown in Fig.~\ref{fig:susceptibility-memory-harmonic-app} and
the time at which the velocity correlator $\Psi(t)$
reaches the first minimum.

\begin{figure}[tb]
\includegraphics[width=0.8\linewidth]{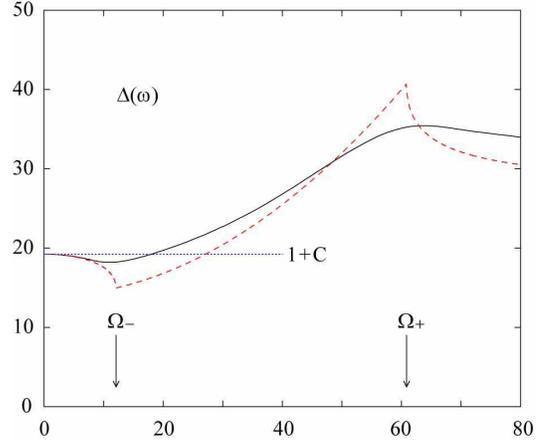}
\caption{
Reactive part of the longitudinal 
modulus $\Delta(\omega)$ (solid line)
for $\rho = 1.093$ and $T = 0.5$ 
taken from Fig.~\ref{fig:Delta-zm-qnull}, 
but plotted here on the linear-$\omega$ axis. 
Dashed line exhibits the SGA modulus
$\Delta_{\rm SGA}(\omega)$ based on 
Eq.~(\ref{eq:Delta-SGA}).
Horizontal dotted line denotes 
$\Delta(\omega = 0) = 1 + C$.
Vertical arrows mark the threshold frequencies $\Omega_{-}$ and
$\Omega_{+}$ defined in the text.}
\label{fig:Delta-harmonic-app}
\end{figure}

There is another interesting
feature caused by the dominant harmonic
nature of the dynamics in stiff-glass states.
From $\Delta(\omega)$ for $T = 0.5$ shown in
Fig.~\ref{fig:Delta-zm-qnull}(b), 
one observes the decrease of $\Delta(\omega)$
from $\Delta(\omega = 0) = 1 + C$
at $\log_{10} \omega \approx 1$.
One can show that such behavior is also 
well described by the harmonic approximation:
it follows from 
Eqs.~(\ref{eq:GLE-hat-phi-c}) and (\ref{eq:harmonic-app-a})
that the reactive part of the modulus within the SGA 
is given by 
\begin{equation}
\Delta_{\rm SGA}(\omega) = (1+C) \, 
\{ \, 1 - w_{1} \, [ \, \tilde{\chi}'(\omega) - 1 \, ] \, \}.
\label{eq:Delta-SGA}
\end{equation}
Figure~\ref{fig:Delta-harmonic-app} compares $\Delta(\omega)$
for $T = 0.5$
with the curve based on Eq.~(\ref{eq:Delta-SGA})
on the linear-$\omega$ axis. 
One understands that 
decrease of $\Delta(\omega)$ from $\Delta(\omega = 0) = 1 + C$
observable at $\omega \approx 10$ 
is due to the harmonic modes near the 
threshold frequency $\Omega_{-}$.
Similarly, the bump at $\omega \approx 60$ reflects
the harmonic modes near $\Omega_{+}$. 

\subsection{Structural-relaxation and vibrational-dynamics
contributions to sound-velocity dispersion}
\label{sec:sound-4}

Here we provide an explanation of the anomalous 
sound-velocity dispersion shown in 
Figs.~\ref{fig:dispersion-liquids}
and \ref{fig:dispersion-glasses} 
based on Eq.~(\ref{eq:positive-dispersion-b})
and on the features of $\Delta(\omega)$ explored in 
the previous two subsections.

At $T = 10$, the whole relaxation of the memory kernel $m(t)$
occurs on the short time scale $\log_{10} t \lesssim -1$
(Fig.~\ref{fig:zm-qnull}).
This yields a susceptibility spectrum $\omega m''(\omega)$
consisting only of the microscopic peak
located at $\log_{10} \omega \approx 1.5$ 
(Fig.~\ref{fig:susceptibility-zm-qnull}),
and the entire increase of $\Delta(\omega)$
occurs in the frequency regime $\log_{10} \omega > 0$
(Fig.~\ref{fig:Delta-zm-qnull}). 
It is therefore sufficient to use the linear-$\omega$
axis to describe the whole $\omega$ variation in 
$\Delta(\omega)$, and this explains via 
Eq.~(\ref{eq:positive-dispersion-b}) 
the observation in 
Fig.~\ref{fig:dispersion-liquids}(a) that
the entire positive dispersion of the sound velocity 
$v_{q}$ starting from $v_{0}$ up to close to $v_{\infty}$
occurs within the linear-$q$ axis.

By decreasing the temperature to $T = 4.0$,
the relaxation of $m(t)$ becomes slower and stretched due to the
development of the cage effect (Fig.~\ref{fig:zm-qnull}), leading
to the evolution of low-frequency contributions in
$\omega m''(\omega)$ and $\Delta(\omega)$
(Figs.~\ref{fig:susceptibility-zm-qnull} and \ref{fig:Delta-zm-qnull}).
Translated to $v_{q}$ via
Eq.~(\ref{eq:positive-dispersion-b}),
this means that the approach of $v_{q}$ towards the hydrodynamic
value $v_{0}$ becomes delayed compared to the one at $T = 10$.
This explains why $v_{q}$ for small wave numbers 
shown in Fig.~\ref{fig:dispersion-liquids}(b)
is still located above $v_{0}$.

\begin{figure}[tb]
\includegraphics[width=0.8\linewidth]{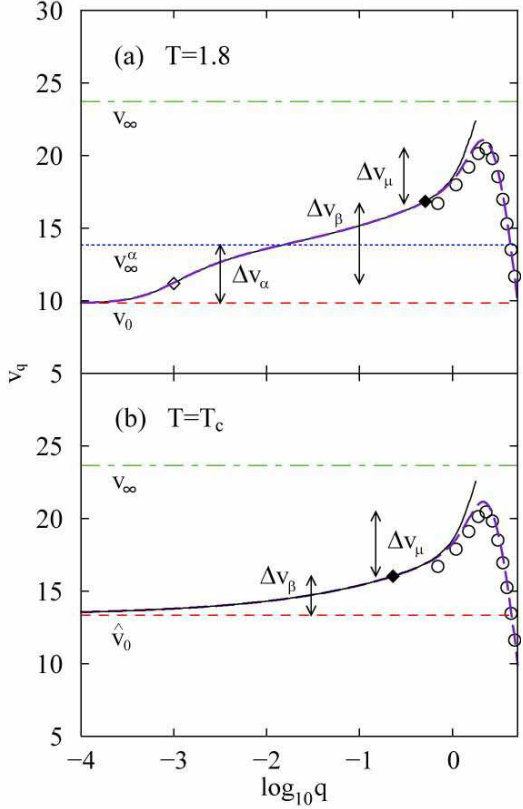}
\caption{
(a) Apparent sound velocity $v_{q} = \omega_{q}^{\rm L \, max}/q$ 
(circles) 
for $\rho = 1.093$ and $T = 1.8$, 
taken from the inset of Fig.~\ref{fig:dispersion-liquids}(c),
but plotted here with the $\log_{10} q$ axis. 
Long-dashed line is from the solution to 
Eq.~(\ref{eq:lowest-order-GHD}),
while solid line is based on Eq.~(\ref{eq:positive-dispersion-b}).
Horizontal lines denote 
$v_{0}$ (dashed line), 
$v_{\infty}$ (dash-dotted line), and
$v_{\infty}^{\alpha}$ (dotted line). 
The contributions $\Delta v_{\alpha}$, $\Delta v_{\beta}$, and
$\Delta v_{\mu}$ to the sound-velocity increase defined in the
text are indicated by vertical arrows.
The open and filled diamonds respectively mark
$v_{\beta}^{-}$ and $v_{\beta}^{+}$ defined in the text.
(b) The same as in (a), but here the sound velocity
refers to $T = T_{\rm c}$, taken from the inset of
Fig.~\ref{fig:dispersion-glasses}(a), and
horizontal dashed line denotes 
$\hat{v}_{0} = v_{0} \sqrt{1 + C}$.}
\label{fig:velocity-dispersion-T180}
\end{figure}

For temperatures close to but above $T_{\rm c}$,
the memory kernel $m(t)$ exhibits
the glassy structural relaxations (Fig.~\ref{fig:zm-qnull}),
leading to low-$\omega$ variations in
$\omega m''(\omega)$ and $\Delta(\omega)$ which
can be adequately described only with the
$\log_{10} \omega$ axis
(Figs.~\ref{fig:susceptibility-zm-qnull} and \ref{fig:Delta-zm-qnull}).
Let us analyze the sound-velocity increase for $T = 1.8$ in detail.
For this purpose, we replot the inset 
of Fig.~\ref{fig:dispersion-liquids}(c)
in Fig.~\ref{fig:velocity-dispersion-T180}(a), 
but with the $\log_{10} q$ axis.
This is necessary to fully appreciate the small-$q$
variation of the sound velocity $v_{q}$, which 
on the basis of Eq.~(\ref{eq:positive-dispersion-b})
originates from the glassy low-$\omega$ 
variations in $\Delta(\omega)$. 
Notice that the circles from the full MCT solutions 
for $\phi_{q}^{{\rm L}''}(\omega)$ can be obtained
only up to the minimum value of the discretized
wave-number grids. 
On the other hand, there is no such restriction
in the solution (solid line) to the lowest-order equation 
(\ref{eq:positive-dispersion-b}) of the 
generalized hydrodynamic description. 
The difference between the circles and the solid line
-- in the $q$ range where the circles are available --
is due to the approximations involved, but 
there should be no problem to use the approximate solution
for understanding the essence of the anomalies.

Let us quantify the contributions to the sound velocity increase
due to the $\alpha$ process 
(to be denoted as $\Delta v_{\alpha}$),
due to the $\beta$ process ($\Delta v_{\beta}$),
and due to the microscopic process ($\Delta v_{\mu}$)
in the following way based on the 
$\alpha$ and $\beta$ regimes for $\Delta(\omega)$
defined at the end of Sec.~\ref{sec:sound-2}. 
As mentioned there, $\Delta(\omega)$ varies 
from $\Delta(\omega = 0) = 1$ to 
$\Delta(\omega_{\alpha}^{+}) = 1 + C^{\rm c}$
in the $\alpha$ regime. 
One therefore obtains from Eq.~(\ref{eq:positive-dispersion-b}):
$\Delta v_{\alpha} = v_{\infty}^{\alpha} - v_{0}$
with $v_{\infty}^{\alpha} = v_{0} \sqrt{1 + C^{\rm c}}$.
The notation comes from the fact that 
$v_{\infty}^{\alpha}$ is the 
high-frequency limit of the sound velocity
in the $\alpha$ regime.
The horizontal dotted line in 
Fig.~\ref{fig:velocity-dispersion-T180}(a)
marks $v_{\infty}^{\alpha}$,
and corresponding linear dispersion law $q v_{\infty}^{\alpha}$ 
and the velocity $v_{\infty}^{\alpha}$ 
have been included with dotted lines in 
Fig.~\ref{fig:dispersion-liquids}(c). 
The $\beta$-relaxation contribution $\Delta v_{\beta}$ 
can be obtained in terms of the frequencies 
$\omega_{\beta}^{\pm}$ also introduced at the end of 
Sec.~\ref{sec:sound-2}:
one obtains from Eq.~(\ref{eq:positive-dispersion-b})
$\Delta v_{\beta} = v_{\beta}^{+} - v_{\beta}^{-}$
with $v_{\beta}^{\pm} = v_{0} \sqrt{\Delta(\omega_{\beta}^{\pm})}$.
These velocities are marked by open
($v_{\beta}^{-}$) and filled ($v_{\beta}^{+}$)
diamonds in Fig.~\ref{fig:velocity-dispersion-T180}(a).
The rest of the sound-velocity increase towards $v_{\infty}$ 
is due to the microscopic process.
As mentioned in the previous subsection,
considerable portion of the microscopic process
is due to the harmonic vibrational dynamics. 
The contributions $\Delta v_{\alpha}$,
$\Delta v_{\beta}$, and $\Delta v_{\mu}$ 
are indicated by vertical arrows in
Fig.~\ref{fig:velocity-dispersion-T180}(a). 
Since there is an overlap between the $\alpha$ and $\beta$ 
regimes ({\em cf.} Sec.~\ref{sec:theory-3}),
there is a corresponding overlap between 
$\Delta v_{\alpha}$ and $\Delta v_{\beta}$. 

The full sound-velocity variation for $T = 1.8$ 
can be understood in this way in terms of the contributions
from $\alpha$- and $\beta$-relaxation processes and from the
microscopic vibrational dynamics. 
Notice that the structural-relaxation contributions 
can be adequately described only with the logarithmic $q$ axis.
This explains why in
Fig.~\ref{fig:dispersion-liquids}(c)
the approach of $v_{q}$ towards 
the hydrodynamic value $v_{0}$
could not be resolved, 
but only the approach towards $v_{\infty}^{\alpha}$ can be
observed. 
Thus, the sound-velocity variation shown
in Fig.~\ref{fig:dispersion-liquids}(c)
mostly reflects the microscopic contribution 
$\Delta v_{\mu}$.

In glass states, the $\alpha$ process is arrested,
and the hydrodynamic sound velocity is altered to
$\hat{v}_{0} = v_{0} \sqrt{1 + C}$.
There are at most $\beta$ and $\mu$ contributions
to the sound-velocity increase from this hydrodynamic value.
At and near $T_{\rm c}$, the $\beta$ contribution,
$\Delta v_{\beta} = v_{\beta}^{+} - \hat{v}_{0}$,
cannot be neglected, where
$v_{\beta}^{+} = v_{0} \sqrt{\Delta(\omega_{\beta}^{+})}$
with $\omega_{\beta}^{+}$ for glasses
defined at the end of Sec.~\ref{sec:sound-2}.
This feature can be understood from
Fig.~\ref{fig:velocity-dispersion-T180}(b),
a redrawing of the inset of 
Fig.~\ref{fig:dispersion-glasses}(a)
with the logarithmic $q$ axis, where
the $\beta$ contribution $\Delta v_{\beta}$
and the microscopic contribution $\Delta v_{\mu}$
(defined as the rest of the sound-velocity increase towards
$v_{\infty}$) are marked by vertical arrows.
The $\beta$-relaxation contribution $\Delta v_{\beta}$ cannot
be resolved with the linear-$q$ axis,
and this explains 
why in Fig.~\ref{fig:dispersion-glasses}(a)
the approach of $v_{q}$ towards $\hat{v}_{0}$ with
decreasing $q$ is retarded. 

For temperatures further away from $T_{\rm c}$, 
the $\beta$-relaxation contribution becomes smaller, 
and the sound-velocity increase from $\hat{v}_{0}$
is entirely due to the microscopic process.
This explains why in Fig.~\ref{fig:dispersion-glasses}(b)
the approach of $v_{q}$ towards $\hat{v}_{0}$
can be observed for $T = 1.47$ even with the linear-$q$ axis. 
By further decreasing the temperature to $T = 0.5$,
a new feature shows up reflecting the dominant
harmonic nature in the stiff-glass-state dynamics.
As discussed in connection with Fig.~\ref{fig:Delta-harmonic-app},
there appears a frequency interval where $\Delta(\omega)$
becomes smaller than $\Delta(\omega = 0)$.  
According to Eq.~(\ref{eq:positive-dispersion-b}),
this gives rise to a ``negative'' dispersion,
and explains the result for $T = 0.5$
shown in Fig.~\ref{fig:dispersion-glasses}(c).

\section{Low-frequency excitations in 
longitudinal and transversal current spectra}
\label{sec:transversal}

The evolution of the AOP is related to the arrest of density
fluctuations caused by the cage effect, and
the AOP is the result of a mapping of the cage distribution 
on the frequency axis ({\em cf.} Sec.~\ref{sec:sound-3}). 
Since the arrest of density fluctuations 
is driven by the ones with a wave number
$q$ near the structure-factor-peak 
position~\cite{Bengtzelius84},
this interpretation of the AOP suggests that it 
appears in spectra of all probing variables 
that couple to density fluctuations of short wavelengths.
But different probing variables will weight the oscillating
complexes differently, and therefore the shape and the peak position
of the AOP will depend somewhat on the probe.
Two such examples have been analyzed in Ref.~\cite{Goetze00}:
the tagged-particle density correlators and the 
velocity correlator $\Psi(t)$.
In particular, it has been shown there how the AOP in 
the memory-kernel spectrum 
$\hat{m}_{s}''(\omega)$
for $\Psi(t)$
({\em cf.} Fig.~\ref{fig:hat-memories-T050}(c))
accounts for the excess intensity in the density of states
$2 \Psi''(\omega) / \pi$ 
with respect to the Debye model.

In this section, we investigate how the AOP manifests itself
in spectral features of the 
longitudinal and transversal current spectra. 
This issue is relevant since in recent computational 
studies~\cite{Horbach01b,Pilla04}
two excitations have been observed
in the longitudinal current spectra,
but the low-frequency boson-peak-like
excitations have been assigned to the transversal mode.
Since there is no direct cross correlation between
the longitudinal and transversal current dynamics, 
the appearance of the transversal mode in the
longitudinal spectra is interpreted as being due to 
``mixing'' phenomena~\cite{Sampoli97}.
This is in contrast to the above interpretation of the AOP,
according to which possible low-frequency excitations in both
of the longitudinal and transversal current spectra should be
the manifestation of the AOP.
Indeed, the AOP is present also in the memory kernel
for the transversal current correlators, and its 
spectral features are quite similar to the one for the
longitudinal current correlators 
as can be inferred from 
Fig.~\ref{fig:hat-memories-T050}(c).
To corroborate our statement, 
we will also study the density dependence of 
spectral features, because such study was claimed to support
the interpretation in terms of the mixing of the longitudinal
and transversal modes
({\em cf.} Sec.~\ref{sec:introduction}).

\subsection{Manifestation of AOP in spectral features}
\label{sec:transversal-1}

\begin{figure}[tb]
\includegraphics[width=0.8\linewidth]{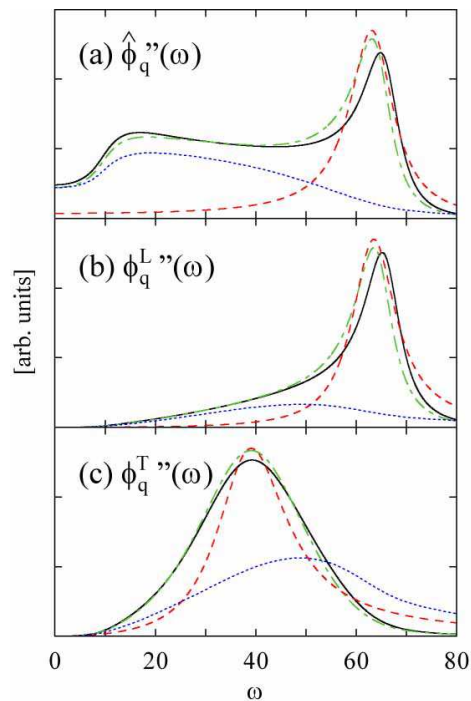}
\caption{
(a) Density fluctuation spectrum $\hat{\phi}_{q}''(\omega)$,
(b) longitudinal current spectrum $\phi_{q}^{{\rm L}''}(\omega)$,
and (c) transversal current spectrum $\phi_{q}^{{\rm T}''}(\omega)$ 
at wave number $q = 3.7$ 
for $\rho = 1.093$ and $T = 0.5$.
Solid lines are based on the full MCT equations of motion
described in Sec.~\ref{sec:theory}.
Dash-dotted lines show curves within the 
generalized hydrodynamic description, Eqs.~(\ref{eq:decomposition-basic}).
Dashed lines denote the HFS part 
based on Eqs.~(\ref{eq:decomposition-HFS}),
whereas dotted lines refer to the AOP portion
from Eqs.~(\ref{eq:decomposition-AOP-1}).
Curves for $\hat{\phi}_{q}''(\omega)$ are obtained
from those for $\phi_{q}^{{\rm L}''}(\omega)$
via Eq.~(\ref{eq:continuity}).}
\label{fig:Sqw-CL-CT}
\end{figure}

We present with solid lines in Fig.~\ref{fig:Sqw-CL-CT}
typical density fluctuation spectrum $\hat{\phi}_{q}''(\omega)$,
longitudinal current spectrum $\phi_{q}^{{\rm L}''}(\omega)$,
and transversal current spectrum $\phi_{q}^{{\rm T}''}(\omega)$
at deep in the glass state, $T = 0.5 \ll T_{c}$. 
These results are based on the full MCT equations of motion
described in Sec.~\ref{sec:theory}. 
The wave number $q=3.7$ chosen for the demonstration is 
about half of the structure-factor-peak position 
({\em cf.} Fig.~\ref{fig:fq-Sq}). 
The fluctuation spectrum $\hat{\phi}_{q}''(\omega)$ 
shown in Fig.~\ref{fig:Sqw-CL-CT}(a) exhibits
two peaks for $\omega \gtrsim 10$, and illustrates a 
hybridization of the high-frequency sound 
with the modes building the AOP~\cite{Goetze00}:
the narrow peak located at higher frequency
is due to the high-frequency sound propagation, and 
a broad lower-frequency excitation reflects the AOP
studied in Sec.~\ref{sec:sound-3}.
On the other hand, there is apparently only one
peak in the longitudinal current spectrum 
$\phi_{q}^{{\rm L}''}(\omega)$ shown in Fig.~\ref{fig:Sqw-CL-CT}(b).
It is due to the high-frequency sound, and the peak 
position $\omega_{q}^{\rm L \, max}$ is nearly the
same as the one for the fluctuation spectrum.
It is not clear how the AOP manifests itself in the
spectral shape of $\phi_{q}^{{\rm L}''}(\omega)$ since,
compared to $\hat{\phi}_{q}''(\omega)$, 
the low frequency part is suppressed due
to the presence of the factor $\omega^{2}$
({\em cf.} Eq.~(\ref{eq:continuity})).
The transversal current spectrum $\phi_{q}^{{\rm T}''}(\omega)$
shown in Fig.~\ref{fig:Sqw-CL-CT}(c) also exhibits only one peak. 
Its peak frequency $\omega_{q}^{\rm T \, max}$
is located at a lower frequency than
$\omega_{q}^{\rm L \, max}$ 
reflecting a smaller transversal sound velocity
$\omega_{q}^{\rm T \, max}/q$
compared to the longitudinal one.
Again, it is not clear whether there is a contribution to
the spectral shape of  
$\phi_{q}^{{\rm T}''}(\omega)$ from the AOP. 
It is the purpose of this subsection to clarify this point.

\begin{figure}[tb]
\includegraphics[width=0.7\linewidth]{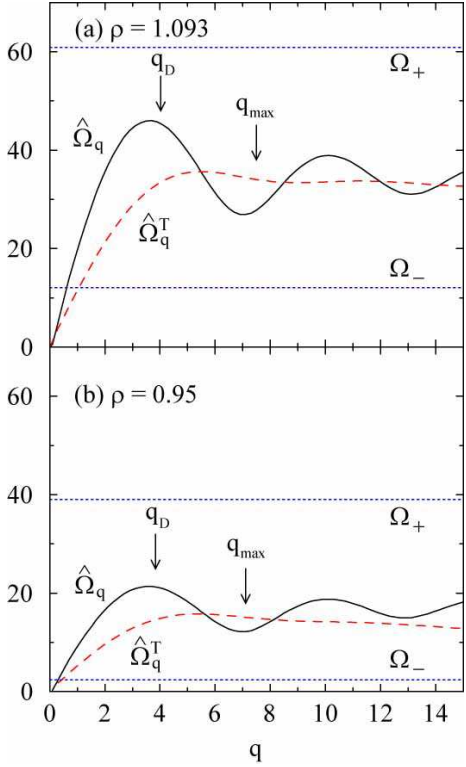}
\caption{
Bare dispersion relation $\hat{\Omega}_{q}$ for the 
longitudinal sound (solid line) and 
$\hat{\Omega}_{q}^{\rm T}$ for the transversal sound
(dashed line) at $T = 0.5$ for
$\rho = 1.093$ (a) and
$\rho = 0.95$ (b).
Horizontal dotted lines denote the threshold frequencies
$\Omega_{\pm}$ 
({\em cf.} Fig.~\ref{fig:hat-memories-T050-2}), and
arrows mark the Debye wave number $q_{\rm D}$ and
the structure-factor-peak position $q_{\rm max}$.} 
\label{fig:hat-Omega}
\end{figure}

\begin{figure}[tb]
\includegraphics[width=0.8\linewidth]{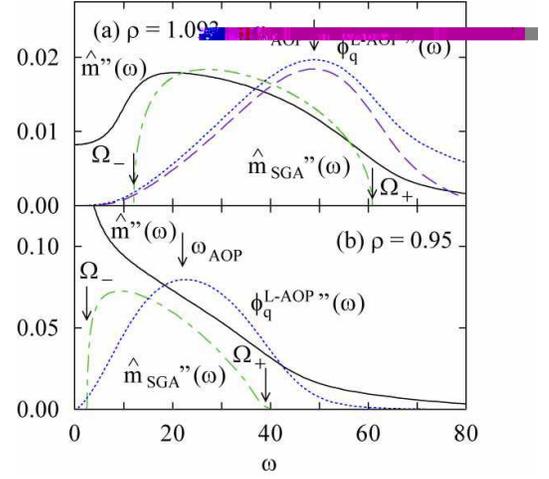}
\caption{
(a) Solid line denotes the memory-kernel spectrum $\hat{m}(\omega)$
for $\rho = 1.093$ and $T = 0.5$, 
while dash-dotted line refers to the SGA spectrum based
on Eqs.~(\ref{eq:harmonic-app}). 
Arrows mark the threshold frequencies $\Omega_{\pm}$
defined below Eqs.~(\ref{eq:harmonic-app}).
Dotted line denotes the frequency dependence of the function
$\phi_{q}^{{\rm L}\mbox{-}{\rm AOP}''}(\omega)$
given in Eq.~(\ref{eq:decomposition-AOP-L-1})
in arbitrary units, and its peak frequency $\omega_{\rm AOP}$
is marked by an arrow.
Long dashed line denotes the corresponding result based on
Eq.~(\ref{eq:decomposition-AOP-L-2}).
(b) The same as in (a), 
but for $\rho = 0.95$~\cite{comment-Thomas-SGA}.}
\label{fig:hat-memories-T050-2}
\end{figure}

To this end, we will derive approximate formulas
specialized to describe each of the high-frequency-sound (HFS)
and low-frequency-AOP portions of the spectra. 
The starting point is the following expression
under the generalized hydrodynamic description 
({\em cf.} Eq.~(\ref{eq:GLE-L-GHD}) and 
citation~\cite{comment-GHD}),
obtained from Eq.~(\ref{eq:GLE-L-glass})
with the hatted memory kernel $\hat{m}_{q}(\omega)$ approximated
by its $q \to 0$ limit,
$\hat{m}(\omega) = \hat{m}_{q=0}(\omega)$: 
\begin{subequations}
\label{eq:decomposition-basic}
\begin{equation}
\phi_{q}^{{\rm L}''}(\omega) =
\frac{\omega^{2} \hat{\Omega}_{q}^{2} \hat{m}''(\omega)}
{[\omega^{2} - \hat{\Omega}_{q}^{2} \hat{\Delta}(\omega)]^{2} +
[\omega \hat{\Omega}_{q}^{2} \hat{m}''(\omega)]^{2}}.
\label{eq:decomposition-basic-1}
\end{equation}
Here $\hat{\Delta}(\omega) = 1 - \omega \hat{m}'(\omega)$. 
Corresponding description for the fluctuation spectrum 
$\hat{\phi}_{q}''(\omega)$ 
is derived via Eq.~(\ref{eq:continuity}).
The formula~(\ref{eq:decomposition-basic-1}) describes 
hybridization of two oscillations:
one is a bare sound with dispersion $\hat{\Omega}_{q}$,
whose wave-number dependence is shown in Fig.~\ref{fig:hat-Omega}(a),
and the other is represented by the AOP spectrum in 
$\hat{m}''(\omega)$.
The AOP region in the spectrum $\hat{m}''(\omega)$
shall be defined as
$\Omega_{-} < \omega < \Omega_{+}$ in terms of the threshold
frequencies $\Omega_{\pm}$ introduced via the SGA spectrum
in Sec.~\ref{sec:sound-3}.
These frequencies are marked in
Fig.~\ref{fig:hat-memories-T050-2}(a). 
The hybridization occurs in the range 
$\Omega_{-} < \hat{\Omega}_{q} < \Omega_{+}$
indicated in Fig.~\ref{fig:hat-Omega}(a) 
where the two oscillations overlap. 
The generalized hydrodynamic description 
for the transversal current can be introduced 
similarly on the basis of 
Eq.~(\ref{eq:GLE-T-glass-b}):
\begin{equation}
\phi_{q}^{{\rm T}''}(\omega) =
\frac{\omega^{2} (\hat{\Omega}_{q}^{\rm T})^{2} 
\hat{m}_{\rm T}''(\omega)}
{[\omega^{2} - (\hat{\Omega}_{q}^{\rm T})^{2} 
\hat{\Delta}_{\rm T}(\omega)]^{2} +
[\omega (\hat{\Omega}_{q}^{\rm T})^{2} 
\hat{m}_{\rm T}''(\omega)]^{2}}. 
\label{eq:decomposition-basic-2}
\end{equation}
\end{subequations}
Here 
$\hat{m}_{\rm T}(\omega) = \hat{m}_{q=0}^{\rm T}(\omega)$
and 
$\hat{\Delta}_{\rm T}(\omega) = 1 - \omega \hat{m}_{\rm T}'(\omega)$. 
Since $\hat{m}''(\omega) \approx \hat{m}_{\rm T}''(\omega)$
as shown in Fig.~\ref{fig:hat-memories-T050}(c), 
one understands that
the hybridization of the bare transversal
sound of dispersion $\hat{\Omega}_{q}^{\rm T}$
with the AOP occurs also in the range
$\Omega_{-} < \hat{\Omega}_{q}^{\rm T} < \Omega_{+}$
indicated in Fig.~\ref{fig:hat-Omega}(a). 
Figure~\ref{fig:Sqw-CL-CT} demonstrates that
the generalized hydrodynamic description
(dash-dotted lines) reproduce main features of 
all spectra fairly well. 
In particular, the subtle hybridization of 
the high-frequency sound and the AOP for $\hat{\phi}_{q}''(\omega)$
is treated semiquantitatively correctly, justifying the use of 
Eqs.~(\ref{eq:decomposition-basic}) as starting equations.

The HFS portion of 
$\phi_{q}^{{\rm L}''}(\omega)$ shall 
be approximated by a Lorenzian (multiplied by $\omega^{2}$)
with the width
determined at the peak frequency $\omega_{q}^{\rm L \, max}$:
\begin{subequations}
\label{eq:decomposition-HFS}
\begin{equation}
\phi_{q}^{{\rm L}\mbox{-}{\rm HFS}''}(\omega) =
\frac{\omega^{2} \hat{\Omega}_{q}^{2} 
\hat{m}''(\omega_{q}^{\rm L \, max})}
{[ \omega^{2} - (\omega_{q}^{\rm L \, max})^{2} ]^{2} +
[ \omega \hat{\Omega}_{q}^{2} 
\hat{m}''(\omega_{q}^{\rm L \, max}) ]^{2}}.
\label{eq:decomposition-HFS-1}
\end{equation}
The dashed curve in Fig.~\ref{fig:Sqw-CL-CT}(b) shows the result
based on this formula, and 
the corresponding HFS curve for the fluctuation spectrum
$\propto \phi_{q}^{{\rm L}\mbox{-}{\rm HFS}''}(\omega) / \omega^{2}$
is included in Fig.~\ref{fig:Sqw-CL-CT}(a).
One understands that the formula~(\ref{eq:decomposition-HFS-1})
well describes the HFS portion of these spectra. 
Thus, the rest of the spectral intensity in the low-frequency
regime should be due to the AOP.

Corresponding expression for 
the HFS part of the transversal current spectrum reads
\begin{equation}
\phi_{q}^{{\rm T}\mbox{-}{\rm HFS}''}(\omega) =
\frac{\omega^{2} (\hat{\Omega}_{q}^{\rm T})^{2} 
\hat{m}_{\rm T}''(\omega_{q}^{\rm T max})}
{[ \omega^{2} - (\omega_{q}^{\rm T max})^{2} ]^{2} +
[ \omega (\hat{\Omega}_{q}^{\rm T})^{2} 
\hat{m}_{\rm T}''(\omega_{q}^{\rm T max}) ]^{2}}.
\label{eq:decomposition-HFS-2}
\end{equation}
\end{subequations}
Figure~\ref{fig:Sqw-CL-CT}(c) shows that 
this formula (dashed line) 
fairly catches the main peak of the spectrum, 
but the predicted width is considerably narrower than
the one of the full MCT solution (solid line). 
This feature can be understood in terms of the AOP 
contribution as follows.
We first notice that $\omega_{q}^{\rm T \, max} \approx 39$ 
for $q = 3.7$ is located in the central region of the AOP 
({\em cf.} Fig.~\ref{fig:hat-memories-T050-2}(a)).
Let us introduces frequencies $\tilde{\omega}_{\pm}$ 
in this central regime so that
$\tilde{\omega}_{-} < \omega_{q}^{\rm T \, max} < \tilde{\omega}_{+}$.
Since there holds
$\hat{m}_{\rm T}''(\omega) \approx \hat{m}''(\omega)$
as shown in Fig.~\ref{fig:hat-memories-T050}(c), 
the Kramers-Kronig relation implies  
$\hat{\Delta}_{\rm T}(\omega) \approx \hat{\Delta}(\omega)$.
One therefore understands from Fig.~\ref{fig:Delta-harmonic-app}
that $\hat{\Delta}_{\rm T}(\omega)$ 
in the central regime of the AOP 
increases with $\omega$, i.e.,
$\hat{\Delta}_{\rm T}(\tilde{\omega}_{-}) <
\hat{\Delta}_{\rm T}(\omega_{q}^{\rm T \, max}) <
\hat{\Delta}_{\rm T}(\tilde{\omega}_{+})$.
Since in lowest order $\omega_{q}^{\rm T \, max}$
is the solution to the equation
$(\omega_{q}^{\rm T \, max})^{2} = (\hat{\Omega}_{q}^{\rm T})^{2}
\Delta_{\rm T}(\omega_{q}^{\rm T \, max})$
({\em cf.} Eq.~(\ref{eq:lowest-order-GHD})),
one has
$(\hat{\Omega}_{q}^{\rm T})^{2} \Delta_{\rm T}(\tilde{\omega}_{-}) <
(\omega_{q}^{\rm T \, max})^{2} <
(\hat{\Omega}_{q}^{\rm T})^{2} \Delta_{\rm T}(\tilde{\omega}_{+})$.
Combined with 
$\tilde{\omega}_{-}^{2} < 
(\omega_{q}^{\rm T \, max})^{2} < \tilde{\omega}_{+}^{2}$,
this leads to an inequality
$[\tilde{\omega}_{\pm}^{2} - (\omega_{q}^{\rm T \, max})]^{2} >
[\tilde{\omega}_{\pm}^{2} - 
(\hat{\Omega}_{q}^{\rm T})^{2} \Delta_{\rm T}(\tilde{\omega}_{\pm})]^{2}$.
We also notice from Fig.~\ref{fig:hat-memories-T050}(c)
that there is only a weak $\omega$ dependence in 
$\hat{m}_{\rm T}''(\omega)$ in the central regime of the AOP.
All this together, one finds on the basis of 
Eqs.~(\ref{eq:decomposition-basic-2}) and 
(\ref{eq:decomposition-HFS-2}):
$\phi_{q}^{{\rm T}''}(\tilde{\omega}_{\pm}) > 
\phi_{q}^{{\rm T}\mbox{-}{\rm HFS}''}(\tilde{\omega}_{\pm})$.
This way, one understands that 
the residual intensity in Fig.~\ref{fig:Sqw-CL-CT}(c) which 
cannot be described as the HFS portion (dashed line) 
is caused by the frequency dependence of
$\hat{\Delta}_{\rm T}(\omega)$ near $\omega_{q}^{\rm T \, max}$, 
which in turn is due to the AOP.
From the mentioned reasoning, 
it is clear that such deviation from the Lorenzian shape
of the spectrum occurs whenever the sound resonance 
frequency is located in the central regime of the AOP.
This explains why such deviation is small for
the longitudinal current spectrum shown in 
Fig.~\ref{fig:Sqw-CL-CT}(b), whose resonance
frequency $\omega_{q}^{\rm L \, max} \approx 65$ exceeds
$\Omega_{+}$ and is not located in the central
regime of the AOP.
Such larger resonance frequency $\omega_{q}^{\rm L \, max}$ 
than expected from 
the bare dispersion $\hat{\Omega}_{q}$ for $q = 3.7$ shown in 
Fig.~\ref{fig:hat-Omega}(a) is due to the 
positive dispersion discussed in Sec.~\ref{sec:sound}.

In view of the results presented so far, one 
understands that there are residual spectral intensities 
due to the AOP in the
frequency regime $\omega < \omega_{q}^{\rm L (T) \, max}$.
In the following, we will derive approximate formulas 
extracting such AOP portion 
based on Eqs.~(\ref{eq:decomposition-basic}).
For this purpose, let us consider the denominator
on the right-hand side of Eq.~(\ref{eq:decomposition-basic-1}),
$[\omega^{2} - \hat{\Omega}_{q}^{2} \, \hat{\Delta}(\omega)]^{2} +
[\omega \, \hat{\Omega}_{q}^{2} \, \hat{m}''(\omega)]^{2}$.
The appearance of the HFS peak 
is mainly due to the decrease of the first term 
towards zero approaching the resonance position 
$\omega_{q}^{\rm L \, max}$.
Since we are interested in the off-resonant regime
$\omega < \omega_{q}^{\rm L \, max}$, 
this term will be approximated by its the small-$\omega$ limit,
$[\omega^{2} - \hat{\Omega}_{q}^{2} \, \hat{\Delta}(\omega)]^{2}
\approx \hat{\Omega}_{q}^{4}$.
Correspondingly, only the leading-order contribution 
for small $\omega$ shall be
retained in the second term, 
$[ \, \omega \, \hat{\Omega}_{q}^{2} \, \hat{m}''(\omega) \, ]^{2} 
\approx [ \, \omega \, \hat{\Omega}_{q}^{2} \, \hat{m}''(\omega=0) \, ]^{2}$.
This results in the following expression for possible AOP portion
in $\phi_{q}^{{\rm L}''}(\omega)$:
\begin{subequations}
\label{eq:decomposition-AOP-1}
\begin{equation}
\phi_{q}^{{\rm L}\mbox{-}{\rm AOP}''}(\omega) =
\frac{1}{\hat{\Omega}_{q}^{2}} 
\frac{\omega^{2} \hat{m}''(\omega)}
{ 1 + [\omega \hat{m}''(\omega = 0)]^{2}}.
\label{eq:decomposition-AOP-L-1}
\end{equation}
According to the derived formula, the $q$ and $\omega$
dependences are factorized.
This implies that the peak position of the AOP portion
-- when it is visible -- is $q$ independent.
The frequency dependence of the 
formula~(\ref{eq:decomposition-AOP-L-1}) is shown as dotted 
line in Fig.~\ref{fig:hat-memories-T050-2}(a), and
its peak frequency, to be denoted as $\omega_{\rm AOP}$, 
is marked by an arrow. 
Furthermore, one infers from the $q$ dependence of 
$\hat{\Omega}_{q}$ shown in Fig.~\ref{fig:hat-Omega} that
the intensity of the AOP portion is predicted to 
increase upon increase of the wave number for
$q \lesssim q_{\rm D}$, 
whereas this trend becomes reversed for
$q \gtrsim q_{\rm D}$. 
Corresponding formula for the 
transversal current spectrum can be derived 
starting from Eq.~(\ref{eq:decomposition-basic-2}):
\begin{equation}
\phi_{q}^{{\rm T}\mbox{-}{\rm AOP}''}(\omega) =
\frac{1}{(\hat{\Omega}_{q}^{\rm T})^{2}} 
\frac{\omega^{2} \hat{m}_{\rm T}''(\omega)}
{ 1 + [\omega \hat{m}_{\rm T}''(\omega = 0)]^{2}}.
\label{eq:decomposition-AOP-T-1}
\end{equation}
\end{subequations}
Since $\hat{m}_{\rm T}''(\omega) \approx \hat{m}''(\omega)$
({\em cf.} Fig.~\ref{fig:hat-memories-T050}(c)),
Eqs.~(\ref{eq:decomposition-AOP-1}) imply that
the peak positions of the AOP portions of both 
longitudinal and transversal current spectra -- 
when they are visible -- 
are located at the same frequency $\omega_{\rm AOP}$.
Furthermore, a stronger intensity of the AOP portion is predicted 
for the transversal current spectra
in the pseudo first Brillouin zone
since there holds $\hat{\Omega}_{q} > \hat{\Omega}_{q}^{\rm T}$
as can be seen from Fig.~\ref{fig:hat-Omega}(a). 
Finally, we notice that 
it is reasonable to further approximate the above formulas as
\begin{subequations}
\label{eq:decomposition-AOP-2}
\begin{eqnarray}
\phi_{q}^{{\rm L}\mbox{-}{\rm AOP}''}(\omega) =
\frac{1}{\hat{\Omega}_{q}^{2}} 
\frac{\omega^{2} \hat{m}^{(1)''}(\omega)}
{ 1 + [\omega \hat{m}^{(2)''}(\omega = 0)]^{2}},
\label{eq:decomposition-AOP-L-2}
\\
\phi_{q}^{{\rm T}\mbox{-}{\rm AOP}''}(\omega) =
\frac{1}{(\hat{\Omega}_{q}^{\rm T})^{2}} 
\frac{\omega^{2} \hat{m}_{\rm T}^{(1)''}(\omega)}
{ 1 + [\omega \hat{m}_{\rm T}^{(2)''}(\omega = 0)]^{2}},
\label{eq:decomposition-AOP-T-2}
\end{eqnarray}
\end{subequations}
since it is the one-mode contribution which provides the
dominant contribution to the memory kernel and 
which describes the AOP, whereas the two-mode contribution
provides a background spectrum 
as discussed in Sec.~\ref{sec:sound-3}.
Figure~\ref{fig:hat-memories-T050-2}(a) demonstrates that
indeed the difference between the curves based on
Eqs.~(\ref{eq:decomposition-AOP-1}) and 
(\ref{eq:decomposition-AOP-2}) is small.
Our motivation for introducing 
Eqs.~(\ref{eq:decomposition-AOP-2}) will become clear
in the next subsection.

\begin{figure}[tb]
\includegraphics[width=0.8\linewidth]{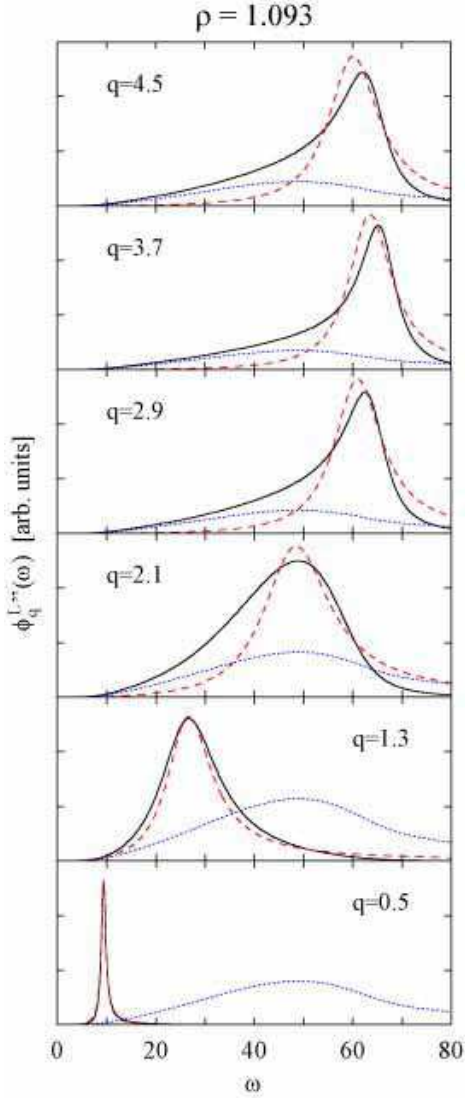}
\caption{
Longitudinal current spectra $\phi_{q}^{{\rm L}''}(\omega)$
(solid lines)
at $\rho = 1.093$ and $T = 0.5$ 
for indicated wave numbers. 
Dashed lines denote the HFS part based on 
Eq.~(\ref{eq:decomposition-HFS-1}), whereas 
dotted lines refer to the AOP portion
obtained from Eq.~(\ref{eq:decomposition-AOP-L-1}).}
\label{fig:CL-1093-T050}
\end{figure}

\begin{figure}[tb]
\includegraphics[width=0.8\linewidth]{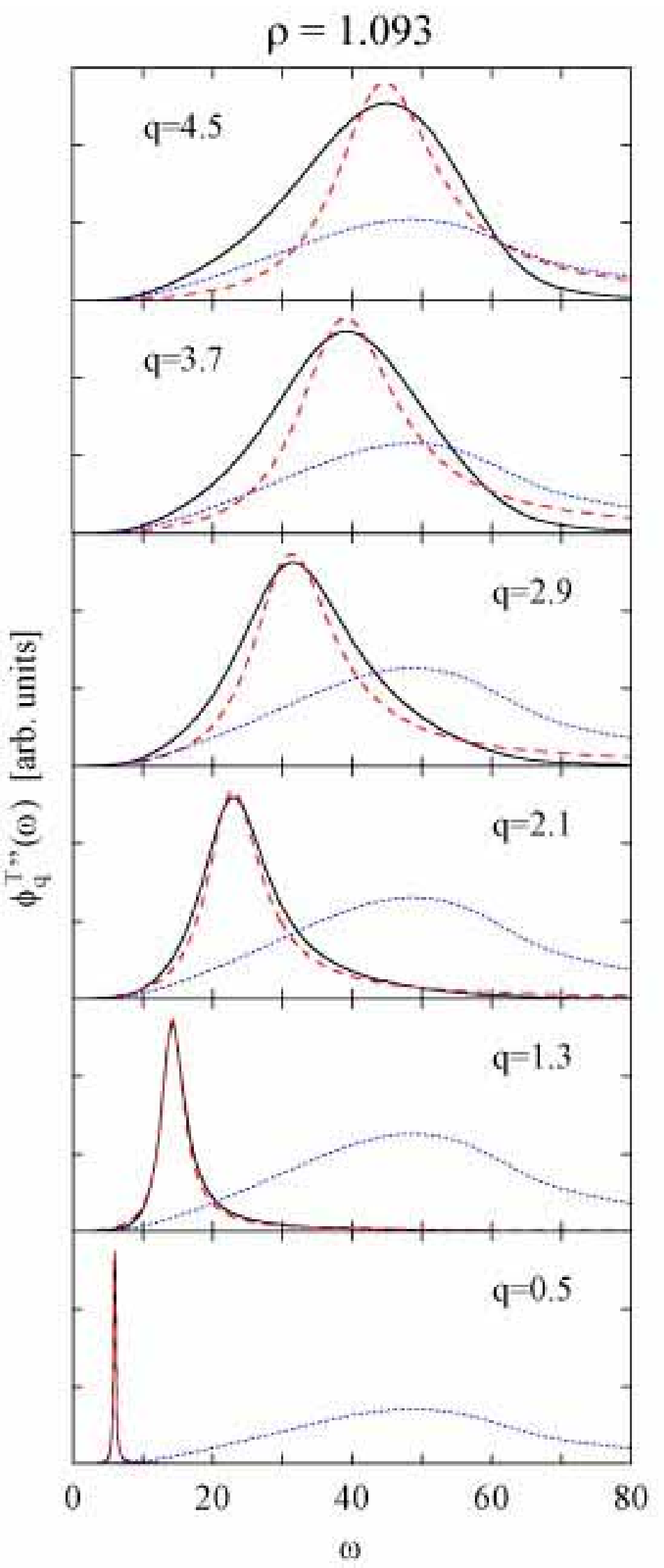}
\caption{
Transversal current spectra $\phi_{q}^{{\rm T}''}(\omega)$
(solid lines)
at $\rho = 1.093$ and $T = 0.5$ 
for indicated wave numbers. 
Dashed lines denote the HFS part based on 
Eq.~(\ref{eq:decomposition-HFS-2}), whereas 
dotted lines refer to the AOP portion
obtained from Eq.~(\ref{eq:decomposition-AOP-T-1}).}
\label{fig:CT-1093-T050}
\end{figure}

\begin{figure}[tb]
\includegraphics[width=0.8\linewidth]{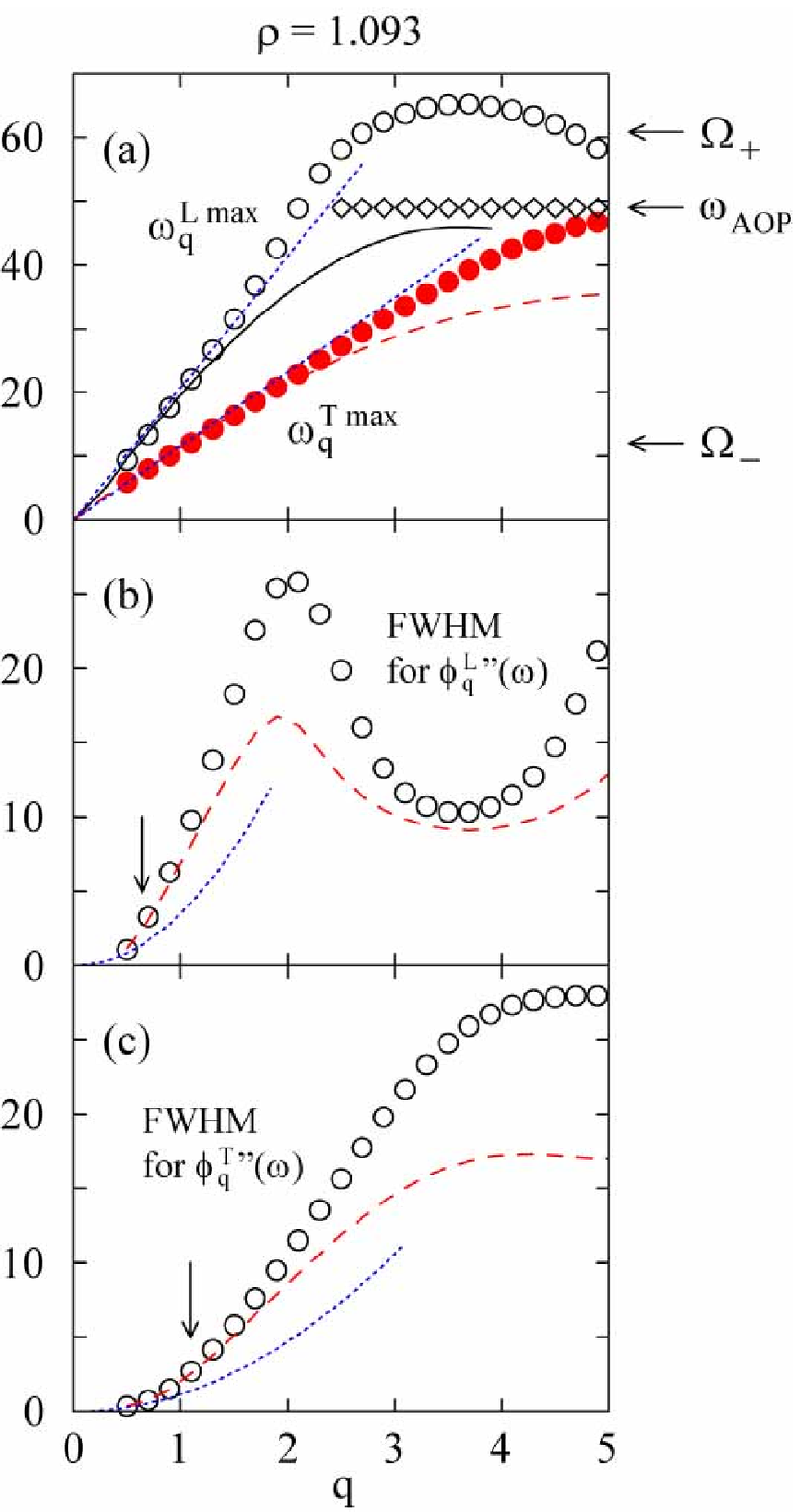}
\caption{
(a) Peak positions $\omega_{q}^{\rm L \, max}$ of 
$\phi_{q}^{{\rm L}''}(\omega)$ (open circles) and 
$\omega_{q}^{\rm T \, max}$ of 
$\phi_{q}^{{\rm T}''}(\omega)$ (filled circles)
as a function of $q$ for $\rho = 1.093$ and $T = 0.5$.
Solid and dashed lines denote the bare dispersions
$\hat{\Omega}_{q}$ and $\hat{\Omega}_{q}^{\rm T}$,
respectively. 
Dotted lines exhibit the hydrodynamic
dispersion laws $q \hat{v}_{0}$ and $q \hat{v}_{0}^{\rm T}$.
Peak positions of the AOP portion
in $\phi_{q}^{{\rm L}''}(\omega)$ 
are denoted by open diamonds.
Horizontal arrows mark $\Omega_{\pm}$ and
$\omega_{\rm AOP}$ taken from
Fig.~\ref{fig:hat-memories-T050-2}(a).
(b) Full width at half maximum (FWHM) for
$\phi_{q}^{{\rm L}''}(\omega)$ (circles) as a function of $q$.
Dashed line denotes the Lorenzian width
$\hat{\Omega}_{q}^{2} \hat{m}''(\omega_{q}^{\rm L \, max})$
predicted by Eq.~(\ref{eq:decomposition-HFS-1}).
Dotted line refers to the hydrodynamic width, 
$q^{2} \hat{v}_{0}^{2} \hat{m}''(\omega=0)$.
Vertical arrow marks the wave number where 
$\omega_{q}^{\rm L \, max} = \Omega_{-}$ holds.
(c) The same as in (b), but for $\phi_{q}^{{\rm T}''}(\omega)$
for which the Lorenzian width (dashed line) is given by
$(\hat{\Omega}_{q}^{\rm T})^{2} \hat{m}_{\rm T}''
(\omega_{q}^{\rm T \, max})$ and 
the hydrodynamic width (dotted line) by
$q^{2} (\hat{v}_{0}^{\rm T})^{2} \hat{m}_{\rm T}''(\omega=0)$.
Vertical arrow marks the wave number where 
$\omega_{q}^{\rm T \, max} = \Omega_{-}$ holds.}
\label{fig:dispersion-T050-rho-1093}
\end{figure}

The dotted lines in Fig.~\ref{fig:Sqw-CL-CT}
are based on Eqs.~(\ref{eq:decomposition-AOP-1}).
For the density fluctuation spectrum (Fig.~\ref{fig:Sqw-CL-CT}(a)), 
it is seen that the dotted line,
obtained from Eq.~(\ref{eq:decomposition-AOP-L-1})
via the relation (\ref{eq:continuity}), 
extracts the AOP portion quite reasonably.
It is also seen from Fig.~\ref{fig:Sqw-CL-CT}(b) that
Eq.~(\ref{eq:decomposition-AOP-L-1}) provides a
reasonable account of the residual low-frequency intensity
in the longitudinal current spectrum
located at $\omega \approx \omega_{\rm AOP}$.
Notice that the AOP appears differently
in $\hat{\phi}_{q}''(\omega)$ and $\phi_{q}^{{\rm L}''}(\omega)$
reflecting the difference of the factor $\omega^{2}$. 
To understand the result shown in Fig.~\ref{fig:Sqw-CL-CT}(c)
for the transversal current spectrum,  
we notice that, by construction, the applicability of the 
formula~(\ref{eq:decomposition-AOP-T-1})
is limited to the small-$\omega$ regime considerably
lower than the sound resonance position $\omega_{q}^{\rm T \, max}$.
Therefore, the dotted line in Fig.~\ref{fig:Sqw-CL-CT}(c)
in the region $\omega \gtrsim \omega_{q}^{\rm T \, max}$ 
has no physical meaning.
Nevertheless, it is seen that the dotted line accounts 
for the residual intensity in the low-frequency regime 
which cannot be described as the HFS portion (dashed line).
Thus, Eq.~(\ref{eq:decomposition-AOP-T-1}) describes the
low-frequency part of the deviation from the Lorenzian 
shape explained above.

The variation of the longitudinal and transversal
current spectra (solid lines)
with changes of wave number $q$ is presented in 
Figs.~\ref{fig:CL-1093-T050} and \ref{fig:CT-1093-T050},
along with their decomposition
into the HFS part (dashed lines) and 
possible AOP portion (dotted lines). 
The dispersion relation and the full width at half maximum
(FWHM) as a function of $q$ are summarized in 
Fig.~\ref{fig:dispersion-T050-rho-1093}.
We start from the discussion on the longitudinal current spectra. 
When $q$ is small so that $\omega_{q}^{\rm L \, max} < \Omega_{-}$,
there is practically no effect from the AOP
on the spectral shape.
(However, the AOP affects the resonance frequency
$\omega_{q}^{\rm L \, max}$ near $\Omega_{-}$
as discussed in Sec.~\ref{sec:sound-4}.) 
Therefore, the spectrum for $q = 0.5$ 
shown in Fig.~\ref{fig:CL-1093-T050}
is dominated by the HFS portion (dashed line),
and there is no physical meaning in the dotted line
for this $q$ value. 
Thus, the spectrum for $q = 0.5$ is very close to Lorenzian
(multiplied by $\omega^{2}$), and its width agrees with
the one from the hydrodynamic prediction 
$q^{2} \hat{v}_{0}^{2} \hat{m}''(\omega=0)$
as demonstrated in Fig.~\ref{fig:dispersion-T050-rho-1093}(b). 
As $q$ increases so that $\omega_{q}^{\rm L \, max}$
becomes larger than $\Omega_{-}$, 
effects from the AOP set in.
These effects increase the spectral width, 
as exemplified for $q = 1.3$ and 2.1 in Fig.~\ref{fig:CL-1093-T050},
in two ways. 
First, the width becomes larger compared to the hydrodynamic width
because 
$\hat{m}''(\omega_{q}^{\rm L \, max}) > \hat{m}''(\omega = 0)$
when the resonance frequency $\omega_{q}^{\rm L \, max}$
is located in the AOP regime, 
$\Omega_{-} < \omega_{q}^{\rm L \, max} < \Omega_{+}$
({\em cf.} Fig.~\ref{fig:hat-memories-T050-2}(a)).
This shows up as the deviation of the dashed line
from the dotted line in Fig.~\ref{fig:dispersion-T050-rho-1093}(b).
In addition, the frequency dependence of $\hat{\Delta}(\omega)$,
which again is caused by the AOP,
also enlarges the width as explained above
in connection with Fig.~\ref{fig:Sqw-CL-CT}(c).
This latter effect leads to a the
non-Lorenzian shape of the spectrum, i.e.,
the deviation of the solid line from the dashed line
for $q = 1.3$ and 2.1 shown in Fig.~\ref{fig:CL-1093-T050},
and also explains the difference between 
corresponding circles and dashed line in 
Fig.~\ref{fig:dispersion-T050-rho-1093}(b). 
The dotted lines in Fig.~\ref{fig:CL-1093-T050}
account for the low-frequency part of the non-Lorenzian 
spectra as discussed above.
As $q$ is increased further, 
the positive dispersion effect becomes important,
and it pushes $\omega_{q}^{\rm L \, max}$ considerably
above frequencies expected from the bare dispersion 
$\hat{\Omega}_{q}$ as shown in
Fig.~\ref{fig:dispersion-T050-rho-1093}(a).
Thereby, there opens a frequency window, where 
$\omega_{\rm AOP} < \omega_{q}^{\rm L \, max}$ holds and 
the appearance of the AOP itself can be observed in the spectra.
This feature holds $q \gtrsim 2.9$ 
as shown in Fig.~\ref{fig:CL-1093-T050}. 
As mentioned above, there is no $q$ dependence in 
$\omega_{\rm AOP}$,
which results in flat dispersion curve 
for the low-frequency excitations as demonstrated with
open diamonds in Fig.~\ref{fig:dispersion-T050-rho-1093}(a). 

Corresponding results for the transversal current spectra
shown in Figs.~\ref{fig:CT-1093-T050} and 
\ref{fig:dispersion-T050-rho-1093} can be explained similarly.
But here, the positive dispersion of $\omega_{q}^{\rm T \, max}$
from the hydrodynamic law
$q \hat{v}_{0}^{\rm T}$ with 
$\hat{v}_{0}^{\rm T} = v_{0} \sqrt{C_{q=0}^{\rm T}}$,
obtained from the leading-order expansion of 
$\hat{\Omega}_{q}^{\rm T}$,
is not so strong as in the longitudinal case.
This is because the bare transversal sound dispersion
$\hat{\Omega}_{q}^{\rm T}$ for $q \lesssim q_{\rm D}$ is 
located below the maximum frequency $\omega \approx 40$ of the
the susceptibility spectrum
$\omega \hat{m}''(\omega)$ 
({\em cf.} the bottom panel for $T = 0.5$ in 
Fig.~\ref{fig:susceptibility-memory-harmonic-app}),
where the positive dispersion effect is most
significant as explained in Sec.~\ref{sec:sound}.
Thus, there is no frequency regime where
$\omega_{\rm AOP} < \omega_{q}^{\rm T \, max}$ holds
for the transversal current spectra, which
can be inferred from 
Fig.~\ref{fig:dispersion-T050-rho-1093}(a). 
Therefore, only the width of $\phi_{q}^{{\rm T}''}(\omega)$
is affected by the AOP, 
which is summarized in Fig.~\ref{fig:dispersion-T050-rho-1093}(c).
However, this feature is altered when the density 
is decreased, as we will see in the next subsection. 

\subsection{Longitudinal and transversal current 
spectra at lower density}
\label{sec:transversal-2}

In this subsection, effects of varying density 
on spectral features of the longitudinal
and transversal current spectra shall be investigated.
Specifically, a density $\rho = 0.95$ lower than 1.093 studied
so far will be considered, but with temperature $T = 0.5$ fixed. 

\begin{figure}[tb]
\includegraphics[width=0.8\linewidth]{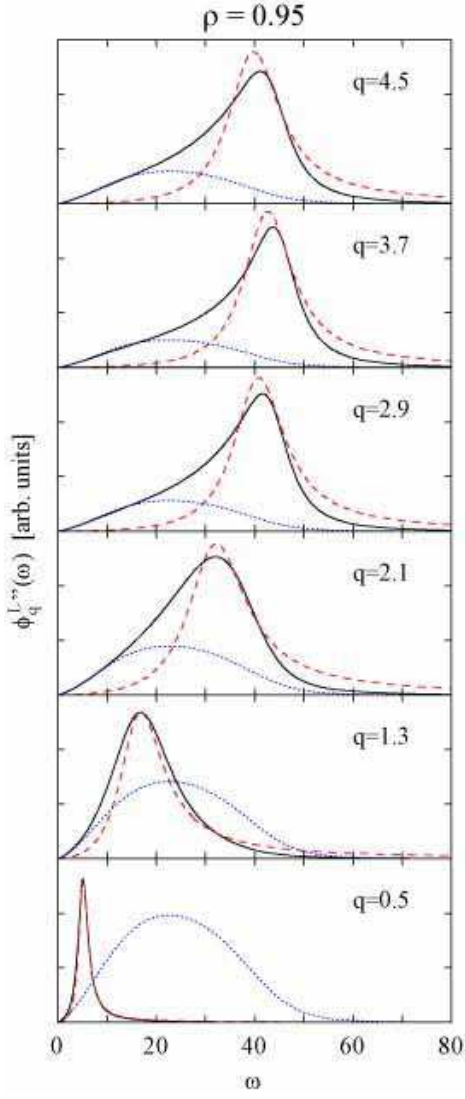}
\caption{
The same as in Fig.~\ref{fig:CL-1093-T050}, but for
a lower density $\rho = 0.95$, and
dotted lines referring to the AOP portion
are based on Eq.~(\ref{eq:decomposition-AOP-L-2})
as discussed in text.}
\label{fig:CL-950-T050}
\end{figure}

\begin{figure}[tb]
\includegraphics[width=0.8\linewidth]{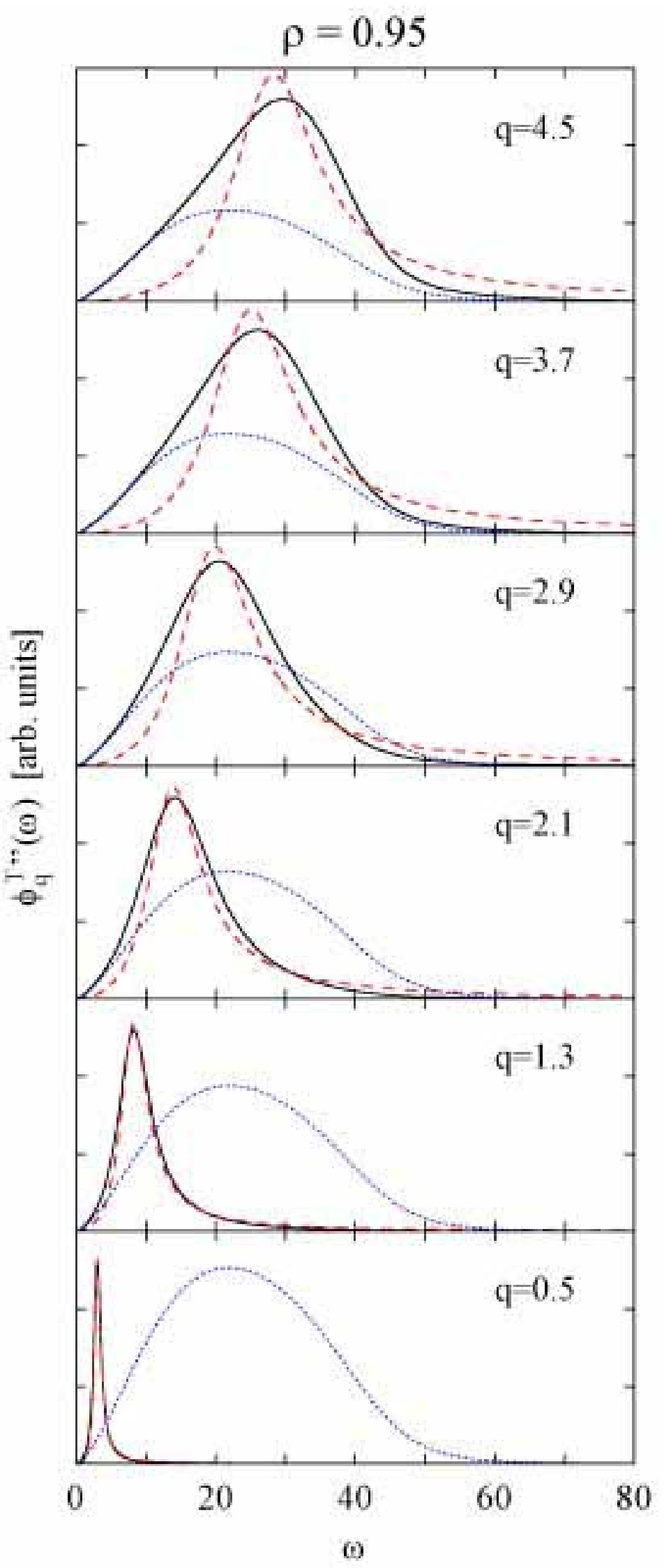}
\caption{
The same as in Fig.~\ref{fig:CT-1093-T050}, but for
a lower density $\rho = 0.95$, and
dotted lines referring to the AOP portion
are based on Eq.~(\ref{eq:decomposition-AOP-T-2})
as discussed in text.}
\label{fig:CT-950-T050}
\end{figure}

\begin{figure}[tb]
\includegraphics[width=0.8\linewidth]{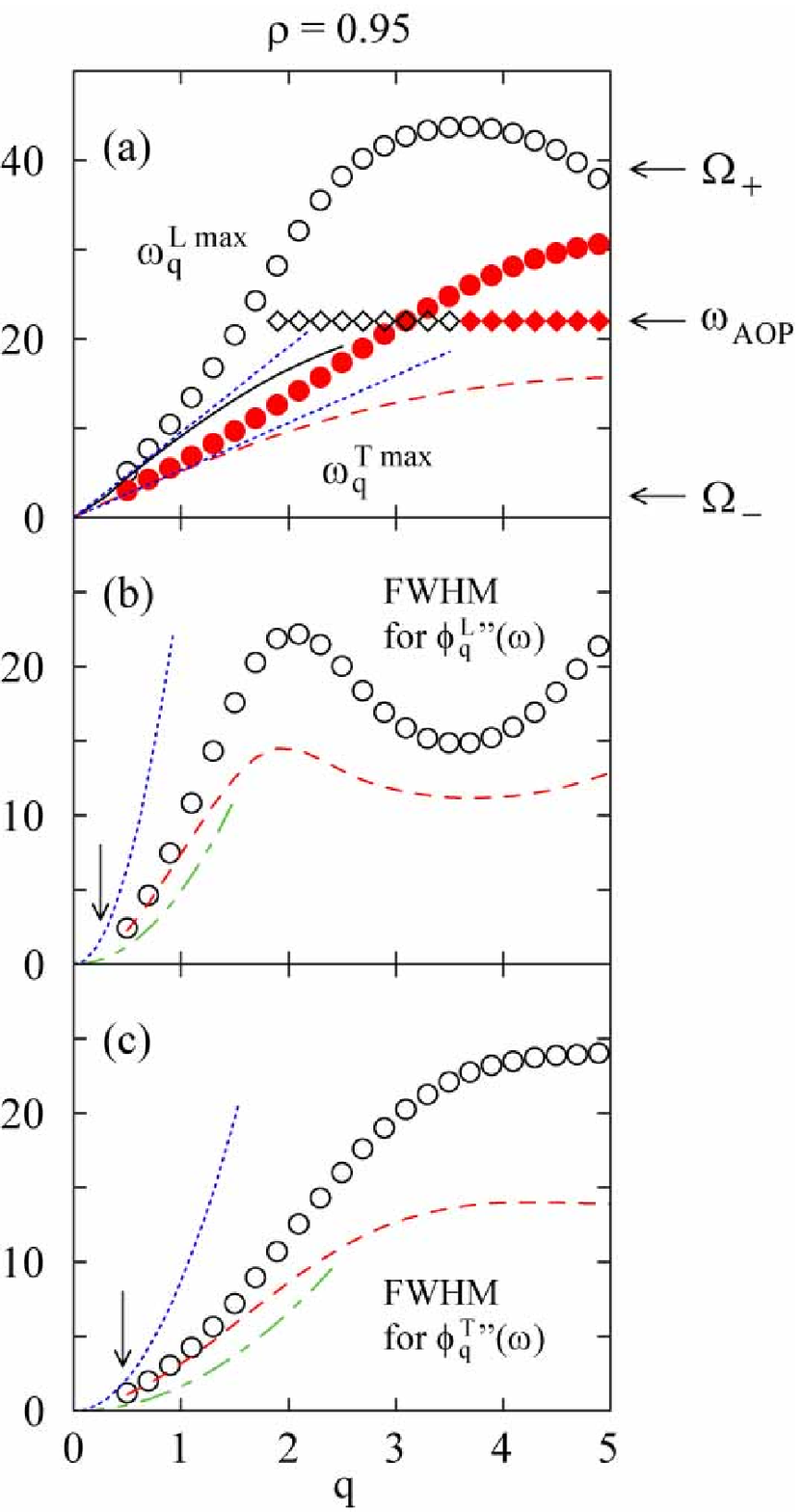}
\caption{
The same as in Fig.~\ref{fig:dispersion-T050-rho-1093},
but for a lower density $\rho = 0.95$.
In (a), filled diamonds are added
denoting the peak positions of the AOP portion
in $\phi_{q}^{{\rm T}''}(\omega)$.
Dash-dotted lines are added in (b) and (c)
denoting $q^{2} \hat{v}_{0}^{2} \hat{m}^{(2)''}(\omega=0)$
and $q^{2} (\hat{v}_{0}^{\rm T})^{2} \hat{m}_{\rm T}^{(2)''}(\omega=0)$,
respectively (see text for details).}
\label{fig:dispersion-T050-rho-950}
\end{figure}

Upon decrease of the density, 
the bare sound dispersions 
$\hat{\Omega}_{q}$ and $\hat{\Omega}_{q}^{\rm T}$ and the
memory-kernel spectrum $\hat{m}''(\omega)$ 
are shifted to lower frequencies as demonstrated
in Figs.~\ref{fig:hat-Omega}(b) and \ref{fig:hat-memories-T050-2}(b),
respectively. 
Notice the striking similarity between the change in 
$\hat{m}''(\omega)$ upon decrease of the density at fixed $T$ 
and the change upon increase of the temperature at fixed $\rho$
found in Fig.~\ref{fig:hat-memories-T050}(b).
Indeed, the critical temperature is decreased to 
$T_{\rm c} \approx 0.644$ for $\rho = 0.95$,
and lowering the density at fixed $T = 0.5$ drives
the system closer to the critical point. 
As discussed with Fig.~\ref{fig:hat-memories-T050}(b),
a strong central peak is formed when the system approaches
the critical point, 
and the modes building the AOP get buried
under its tail. 
But again, Fig.~\ref{fig:hat-memories-T050-2}(b) demonstrates that
the SGA kernel $\hat{m}_{\rm SGA}''(\omega)$ extracts 
the buried AOP portion quite reasonably~\cite{comment-Thomas-SGA}.
Thus, in analogy to what is found in 
Fig.~\ref{fig:hat-memories-T050}(b), 
the AOP shifts towards lower frequencies upon 
decrease of the density, accompanied by an increase of its 
spectral intensity.

As a related problem, we found that the use of 
Eqs.~(\ref{eq:decomposition-AOP-1}) to extract
the AOP portion from current spectra
leads to unplausible results.
This is because of the mentioned development of the central peak, 
leading to quite large value of $\hat{m}''(\omega=0)$.
As will be demonstrated in the following
({\em cf.} Figs.~\ref{fig:dispersion-T050-rho-950}(b) and
\ref{fig:dispersion-T050-rho-950}(c)),
such large value of $\hat{m}''(\omega=0)$ caused by 
the critical decay near $T_{\rm c}$ is not appropriate
for handling the high-frequency dynamics.
We will therefore use Eqs.~(\ref{eq:decomposition-AOP-2}) 
instead of Eqs.~(\ref{eq:decomposition-AOP-1}), 
which essentially amounts to replacing 
$\hat{m}''(\omega=0)$ with $\hat{m}^{(2)''}(\omega=0)$. 
The curves based on Eqs.~(\ref{eq:decomposition-AOP-2})
are not much affected by the critical dynamics because
(i) only corrections to the critical decay enter into
$\hat{m}^{(2)''}(\omega)$ as explained in connection with
Fig.~\ref{fig:hat-memories-T050}(b),
and (ii) the critical-decay contribution
in $\hat{m}^{(1)''}(\omega)$
is suppressed in Eq.~(\ref{eq:decomposition-AOP-L-2})
due to the presence of the factor $\omega^{2}$
in its denominator. 
As demonstrated in Fig.~\ref{fig:hat-memories-T050-2}(a),
there is practically no difference between 
the use of Eqs.~(\ref{eq:decomposition-AOP-1}) and
(\ref{eq:decomposition-AOP-2}) for $\rho = 1.093$,
and we expect that physics is not altered with the use
of the latter for $\rho = 0.95$. 

Longitudinal and transversal
current spectra for $\rho = 0.95$
(solid lines) with changes of wave number $q$ 
are presented in 
Figs.~\ref{fig:CL-950-T050} and \ref{fig:CT-950-T050},
along with their decomposition
into the HFS part (dashed lines) and 
possible AOP portion (dotted lines). 
As mentioned above, the dotted lines in these figures
are based on Eqs.~(\ref{eq:decomposition-AOP-2}).
The dispersion relation and the FWHM as a 
function of $q$ are summarized in 
Fig.~\ref{fig:dispersion-T050-rho-950}.
It is seen from Figs.~\ref{fig:dispersion-T050-rho-950}(b)
and \ref{fig:dispersion-T050-rho-950}(c) that
indeed the hydrodynamic prediction for the width
with $\hat{m}''(\omega = 0)$ does not work for $\rho = 0.95$;
instead, the corresponding prediction with
$\hat{m}^{(2)''}(\omega = 0)$ reasonably accounts for the
width in the small-$q$ regime. 

It is seen by comparing 
Figs.~\ref{fig:CL-1093-T050} and 
\ref{fig:CL-950-T050} that
the AOP portion of the longitudinal current spectra
for the smaller density $\rho = 0.95$ 
is located at lower frequencies and its intensity 
is more enhanced.
These differences simply reflect the shift of the AOP to 
lower frequencies accompanied by an increase of its
spectral intensity for $\rho = 0.95$. 
In addition, the positive dispersion of the sound velocity
and the deviation from the Lorenzian shape of the
spectra start at smaller $q$ for $\rho = 0.95$,
as can be inferred by comparing 
Figs.~\ref{fig:dispersion-T050-rho-1093} and
\ref{fig:dispersion-T050-rho-950}.
These features are also due to the shift of the AOP 
to lower frequencies upon decrease of the density.
Furthermore, there is no ``negative'' dispersion
({\em cf.} Sec.~\ref{sec:sound-4})
for $\rho = 0.95$ since the memory kernel for $T = 0.5$
at this density is not dominated by the harmonic contributions
({\em cf.} Fig.~\ref{fig:hat-memories-T050-2}(b)).
Otherwise, the variation of the longitudinal current spectra 
with changes of $q$ for $\rho = 0.95$ is quite 
similar to the one found for $\rho = 1.093$, 
and the discussion shall not be repeated. 

The variation of the transversal current spectra 
with changes of $q$ for $\rho = 0.95$ is also quite 
similar to the one found for $\rho = 1.093$, 
as can be understood by comparing
Figs.~\ref{fig:CT-1093-T050} and
\ref{fig:CT-950-T050}.
But here,
the resonance frequency $\omega_{q}^{\rm T \, max}$ of the
transversal sound exhibits a strong positive dispersion
as shown in Fig.~\ref{fig:dispersion-T050-rho-950}(a).
Again, this is caused by the shift of the AOP to 
lower frequencies. 
The positive dispersion of the transversal 
sound velocity due to the contribution from the AOP
can be discussed
quite similarly to what is presented in Sec.~\ref{sec:sound}
for the longitudinal sound. 
Thereby, there opens a frequency window also for
the transversal current spectra, where 
$\omega_{\rm AOP} < \omega_{q}^{\rm T \, max}$ holds and
the appearance of the AOP itself can be observed in the spectra.
Such AOP portion can be observed for
$q \gtrsim 3.7$ in Fig.~\ref{fig:CT-950-T050},
yielding a flat dispersion curve 
for the low-frequency excitations of the transversal
current spectra as demonstrated with
filled diamonds in Fig.~\ref{fig:dispersion-T050-rho-950}(a). 
This is in contrast to the result found for $\rho = 1.093$,
where $\omega_{\rm AOP} > \omega_{q}^{\rm T \, max}$ holds
for the whole $q$ range 
and the effects from the AOP showed up only in the 
width of the transversal current spectra.

\section{Concluding remarks}
\label{sec:conclusions}

In this paper, we presented a theoretical investigation
on the high-frequency collective dynamics in liquids
and glasses at microscopic length and time scales
based on the MCT for ideal liquid-glass transition.
We focused on recently investigated issues from 
IXS and computer-simulation studies for 
dynamic structure factors and 
longitudinal and transversal current spectra:
the anomalous dispersion of the high-frequency
sound velocity and the nature of the low-frequency excitation
called the boson peak.
The results of the theory are demonstrated for 
a model of the LJ system. 

The MCT explains the evolution of the structural-relaxation
processes as precursor of the glass transition at a
critical temperature $T_{\rm c}$,
which is driven by the mutual blocking of a particle and its
neighbors (cage effect)~\cite{Bengtzelius84,Goetze91b}.
On the other hand, the theory also predicts the
development of the harmonic vibrational excitations
in stiff-glass states $T \ll T_{\rm c}$, 
called the AOP.
The AOP is caused by the strong interaction between density
fluctuations at microscopic length scales and the 
arrested glass structure, and exhibits the properties
of the boson peak~\cite{Goetze00}.
As demonstrated in Sec.~\ref{sec:sound-3}, the AOP
persists also near and above $T_{\rm c}$, and
considerable portion of the microscopic process
in systems ranging from high-$T$ liquids
down to deep-in-glass states 
can be described as a superposition of harmonic 
vibrational dynamics. 
We investigated in Sec.~\ref{sec:sound} 
how the interference of the sound mode with
these structural-relaxation processes and vibrational
excitations shows up as anomalies in the
sound-velocity dispersion. 

In the vicinity of $T_{\rm c}$,
the structural-relaxation contributions 
to the sound-velocity dispersion show up at very 
small wave numbers, which can fully be appreciated
only with the logarithmic $q$ axis 
({\em cf.} Fig.~\ref{fig:velocity-dispersion-T180}). 
Therefore, with the linear-$q$ axis adopted in 
conventional studies on the dispersion relation, 
the sound velocity $v_{q}$ in the small-$q$ limit
is located above the hydrodynamic value $v_{0}$
({\em cf.} Fig.~\ref{fig:dispersion-liquids}(c)),
and this even at $T = 4.0$ 
({\em cf.} Fig.~\ref{fig:dispersion-liquids}(b))
which is more than twice of the critical 
temperature $T_{\rm c}$.
We notice that this temperature is also higher than
the freezing temperature $T_{\rm f} \approx 3.3$
which is estimated based on the so-called
Hansen-Verlet criterion~\cite{Hansen69}.
Thus, except for very high temperature like $T = 10$
({\em cf.} Fig.~\ref{fig:dispersion-liquids}(a)),
the variation of the sound velocity $v_{q}$ 
which is visible in the linear $q$ axis basically
reflects the contribution from the microscopic process.
This theoretical result is in contrast to the traditional
picture based on the viscoelastic model according to which
the anomalous dispersion is fully ascribed to the
structural relaxation, but is in agreement with finding
from recent IXS studies on simple liquid metals near the
melting temperature~\cite{Scopigno00,Scopigno00b,Scopigno00c,Scopigno02b,Monaco04}.

The description of the microscopic process 
in terms of the AOP -- a superposition of harmonic 
oscillations of particles inside their cages --
implies that the decay of the memory kernel
at microscopic times $\log_{10} t \lesssim -1$
({\em cf.} Fig.~\ref{fig:zm-qnull})
reflects the dephasing of different oscillatory components
in the force fluctuations.
This is in consistent with
the implication from the computer-simulation study
on harmonic glass~\cite{Ruocco00}.
Such microscopic process, despite arrested
structural relaxations, accounts for the 
positive dispersion in glass states
({\em cf.} Fig.~\ref{fig:dispersion-glasses}),
whose presence has been reported from recent MD and IXS 
studies~\cite{Taraskin97,Ruocco00,Horbach01b,Scopigno02,Pilla04,Ruzicka04}.
That the microscopic process due to the AOP persists
both in liquid and glass states
with only weak temperature dependence of its characteristic time
or frequency
({\em cf.} Fig.~\ref{fig:susceptibility-memory-harmonic-app})
is in accord with the computer-simulation 
result presented in Ref.~\cite{Scopigno02}.

\begin{figure}[tb]
\includegraphics[width=0.8\linewidth]{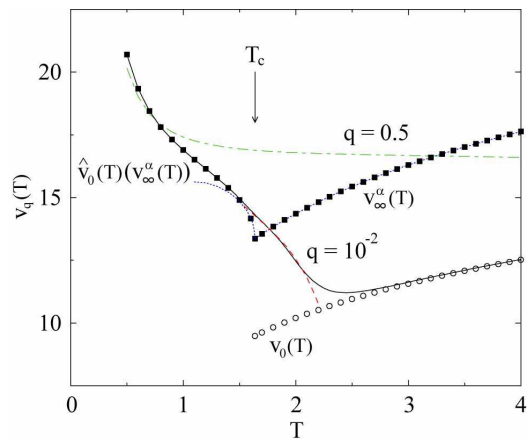}
\caption{
Sound velocities $v_{q}(T)$ as a function of $T$
at $\rho = 1.093$ 
for $q = 10^{-2}$ (solid line) and $q = 0.5$ (dash-dotted line),
which are based on 
Eq.~(\ref{eq:positive-dispersion-b}) 
with $\Delta(\omega)$
determined for each temperature 
({\em cf.} Fig.~\ref{fig:Delta-zm-qnull}). 
Hydrodynamic sound velocities $v_{0}(T)$ for $T > T_{\rm c}$
and $\hat{v}_{0}(T) = v_{\infty}^{\alpha}(T)$ (see text)
for $T \le T_{\rm c}$ are denoted 
as open circles and filled squares, respectively.
$v_{\infty}^{\alpha}(T)$ for $T > T_{\rm c}$, 
given by 
$v_{\infty}^{\alpha}(T) = v_{0}(T) / \sqrt{1 - f_{0}^{\rm c}}$,
are also denoted as filled squares.
Dotted line shows the
square-root singularity translated to $v_{\infty}^{\alpha}(T)$,
i.e.,
$v_{\infty}^{\alpha}(T) = v_{0}(T) \sqrt{1 - f_{0}(T)}$
with
$f_{0}(T) = f_{0}^{\rm c} + h_{0} \sqrt{ \sigma / (1 - \lambda) }$
for $T \le T_{c}$ and
$f_{0}(T) = f_{0}^{\rm c}$ for $T > T_{\rm c}$.
Dashed line exhibits the $\beta$-relaxation contribution to 
$v_{q}(T)$ which is based on Eq.~(\ref{eq:positive-dispersion-b}) 
with $\Delta(\omega)$
determined from the MCT asymptotic formula (\ref{eq:Delta-beta}).}
\label{fig:vq-vs-T}
\end{figure}

To see further implications of the theory,
we show in Fig.~\ref{fig:vq-vs-T}
the sound velocities $v_{q}(T)$ as a function of $T$ 
for fixed $q = 10^{-2}$ (solid line) and $q = 0.5$ (dash-dotted line).
(The argument $T$ shall be added here to 
emphasize the $T$ dependence.)
The density is fixed to $\rho = 1.093$, and
$v_{q}(T)$ are obtained from
Eq.~(\ref{eq:positive-dispersion-b}) 
with $\Delta(\omega)$
determined for each temperature 
({\em cf.} Fig.~\ref{fig:Delta-zm-qnull}). 
Assuming that LJ particles under study
are of argon size ($\approx 0.34$ nm),
the wave number $q = 0.5$ (1.5 nm$^{-1}$) is in the lowest
momentum-transfer range in IXS measurements, and
$q = 10^{-2}$ (0.03 nm$^{-1}$) is a typical
momentum transfer in Brillouin light scattering (BLS)
experiments. 
As shown in Fig.~\ref{fig:dispersion-glasses}(c)
for a stiff-glass state $T = 0.5$, a negative dispersion is observable
for $q < 1$ where the sound velocity $v_{q}(T)$ is
smaller than the hydrodynamic ($q \to 0$) value.
This is due to the dominant harmonic nature of the
dynamics at $T = 0.5$, because of which
there appears a frequency interval where 
$\Delta(\omega)$ is smaller than its $\omega \to 0$ limit
({\em cf.} Fig.~\ref{fig:Delta-harmonic-app}).
The presence of the negative dispersion 
is reflected in Fig.~\ref{fig:vq-vs-T}
as the emergence of the low-$T$ region where 
the IXS sound velocity ($v_{q}(T)$ for $q = 0.5$) is smaller than
the BLS sound velocity ($q = 10^{-2}$).
Such $T$ dependence
of the IXS and BLS sound-velocity data in the low-$T$ region 
has been reported for 
glycerol~\cite{Sette98,Masciovecchio98b}
(see Fig.~5(b) of Ref.~\cite{Sette98}).
Thus, there is an experimental result
which is consistent with the presence
of the negative dispersion predicted by the theory.
That the difference between the IXS and BLS sound-velocity
data in the low-$T$ region is smaller in Fig.~\ref{fig:vq-vs-T} 
than the one found in Fig.~5(b) of Ref.~\cite{Sette98} for glycerol
and that such difference is not clearly observable in 
orthoterphenyl~\cite{Monaco98}
might be rationalized by the 
stronger influence of vibrational modes (boson peaks) 
in the dynamics of 
network glass former glycerol
than in fragile glass formers like the
LJ system and orthoterphenyl~\cite{Sokolov93b}.
According to this reasoning, 
the presence of the negative dispersion is 
also expected in strong glass formers like silica 
where the influence of vibrational dynamics is
even stronger~\cite{Sokolov93b}.
The sound velocity data for vitreous silica reported in Fig.~3 of 
Ref.~\cite{Ruzicka04} seems to indicate 
this possibility,
but large error bars do not allow ones to draw
a decisive conclusion.
It would therefore be of value to re-analyze
the data for silica with improved resolution.

Hydrodynamic sound velocities $v_{0}(T)$ for 
$T> T_{\rm c}$ and $\hat{v}_{0}(T) = v_{0}(T) / \sqrt{1 - f_{0}(T)}$ 
for $T \le T_{\rm c}$ are
included in Fig.~\ref{fig:vq-vs-T} 
as open circles and filled squares, respectively.
$v_{0}(T)$ increases with increasing $T$ because in our study
$T$ is varied with density fixed, leading to
increased pressures at elevated temperatures. 
On the other hand, the $T$ dependence of 
$\hat{v}_{0}(T)$ is dominated by that of $f_{0}(T)$ 
({\em cf.} Fig.~\ref{fig:fq-Sq}),
and this explains the increase of $\hat{v}_{0}(T)$ 
with decreasing $T$. 
Here a side remark shall be added concerning the
notion of the ``hydrodynamic'' sound velocity 
$\hat{v}_{0}(T)$
for glass states referring to $T \le T_{\rm c}$.
The MCT in its idealized form, adopted in the
present study, predicts the structural
arrest at the critical temperature $T_{\rm c}$
which is located above the calorimetric glass-transition
temperature $T_{\rm g}$.
In reality there are slow dynamical processes 
-- referred to as hopping processes --
which restore ergodicity for $T \le T_{\rm c}$.
These processes, which are studied within the
extended version of the MCT~\cite{Goetze87}, 
change the ideal elastic peaks 
$\pi f_{q}(T) \delta(\omega)$ of density fluctuation spectra
$\phi_{q}''(\omega)$ 
into quasielastic $\alpha$ peaks of nonzero width also for
$T \le T_{\rm c}$.
$f_{q}(T)$ in real systems therefore has to be interpreted as 
an effective Debye-Waller factor~\cite{Goetze92}, 
which can be measured as area under the quasielastic $\alpha$ peak
in $\phi_{q}''(\omega)$ or by determining the
plateau of the $\phi_{q}(t)$-versus-$\log t$ curve. 
Correspondingly, $\hat{v}_{0}(T)$ should be interpreted
as the extension of $v_{\infty}^{\alpha}(T)$ 
-- the sound velocity in the high-frequency limit 
of the $\alpha$ regime --
introduced in Sec.~\ref{sec:sound-4} to $T \le T_{\rm c}$.
$v_{\infty}^{\alpha}(T)$ for $T > T_{\rm c}$
are given by 
$v_{\infty}^{\alpha}(T) = v_{0}(T) / \sqrt{1 - f_{0}^{\rm c}}$
({\em cf.} Sec.~\ref{sec:sound-4}),
and are also included with filled squares in Fig.~\ref{fig:vq-vs-T}:
$v_{\infty}^{\alpha}(T)$ defined below and above $T_{\rm c}$
merge at $T_{\rm c}$. 

The mentioned dependence of $v_{\infty}^{\alpha}(T)$ on $f_{0}(T)$
has been utilized 
to extract experimentally the zero wave-number limit of the
Debye-Waller factor via the relation 
$f_{0}(T) = 1 - [v_{0}(T)/v_{\infty}^{\alpha}(T)]^{2}$~\cite{Fuchs90,Li93}
and to test the square-root singularity as predicted by the MCT, 
$f_{0}(T) = f_{0}^{\rm c} + h_{0} \sqrt{ \sigma / (1 - \lambda) }$
for $T \le T_{c}$ and
$f_{0}(T) = f_{0}^{\rm c}$ for $T > T_{\rm c}$
({\em cf.} Eq.~(\ref{eq:plateau-X})).
The dotted line in Fig.~\ref{fig:vq-vs-T} shows the
square-root singularity translated to $v_{\infty}^{\alpha}(T)$.
In our system, there is no possibility to detect the 
singularity based on 
the IXS sound velocity ($v_{q}(T)$ for $q = 0.5$),
which basically reflects the contribution from the microscopic
process
({\em cf.} Fig.~\ref{fig:velocity-dispersion-T180}).
On the other hand, the BLS sound velocity ($q = 10^{-2}$)
probes much lower momentum-transfer range,
and is thus affected also by the structural-relaxation processes.
In particular, the $\beta$-relaxation contribution
to $v_{q}(T)$ plays an important role in the 
vicinity of $T_{\rm c}$ 
({\em cf.} Fig.~\ref{fig:velocity-dispersion-T180}),
as can also be inferred
by comparing the solid and dashed lines in 
Fig.~\ref{fig:vq-vs-T},
the latter being obtained based on 
Eq.~(\ref{eq:positive-dispersion-b}) 
with $\Delta(\omega)$
determined from the MCT asymptotic formula (\ref{eq:Delta-beta})
for the $\beta$ relaxation. 
Thus, the $\beta$ relaxation must be included 
for reliable determination of $f_{0}(T)$
via the sound velocity data, which is in full agreement
with the conclusion drawn in Ref.~\cite{Li93}.
({\em cf.} Fig.~13 of Ref.~\cite{Li93} where
BLS data for CaKNO$_{3}$ 
corresponding to solid line and filled squares in
Fig.~\ref{fig:vq-vs-T} are shown.)

We studied in Sec.~\ref{sec:transversal}
how the AOP manifests itself in the
spectral shape of the glass-state 
longitudinal and transversal current spectra at low frequencies.
For small wave numbers where the sound resonance frequencies
are located below the low-frequency threshold of the AOP, 
$\omega_{q}^{\rm L(T) \, max} < \Omega_{-}$,
there is practically no effect from the AOP, and 
the longitudinal and transversal current spectra
are dominated by the sound excitations of the
Lorenzian spectral shape.
By increasing the wave numbers so that
the sound resonance frequencies enter the AOP region,
$\Omega_{-} < \omega_{q}^{\rm L(T) \, max}$, 
the hybridization of the sound mode with the AOP becomes
important, and the shape of the current 
spectra deviates from Lorenzian. 
As the wave number is increased further, 
the interference of the sound mode
with the AOP leads to the positive dispersion of the
sound velocities,
pushing $\omega_{q}^{\rm L(T) \, max}$ considerably
above frequencies expected from the bare dispersion.
Thereby, there opens a frequency window,
where the appearance of the AOP itself can be observed
in the spectra at low frequencies. 
Furthermore, our approximate formulas~(\ref{eq:decomposition-AOP-1})
predict that (i) there is no $q$ dependence in the peak frequency
$\omega_{\rm AOP}$ of the AOP portions in the
current spectra,
(ii) $\omega_{\rm AOP}$ is nearly the same for both of the
longitudinal and transversal current spectra
since $\hat{m}''(\omega) \approx \hat{m}_{\rm T}''(\omega)$
as shown in Fig.~\ref{fig:hat-memories-T050}(c),
and (iii) the intensity of the AOP portion is stronger
for the transversal current spectra than for the longitudinal
ones in the pseudo first Brillouin zone.
These theoretical predictions, demonstrated in 
Figs.~\ref{fig:CL-1093-T050} to \ref{fig:dispersion-T050-rho-950}
for the LJ system, 
are in agreement with previous findings 
from MD and IXS studies~\cite{Horbach01b,Pilla04,Scopigno03,Ruzicka04}.
Thus, the low-frequency excitations 
observable at the same resonance frequency $\omega_{\rm AOP}$
in both of the glass-state longitudinal and transversal
current spectra are the manifestation of the AOP.
Usually, the boson peaks are discussed below $T_{\rm g}$.
On the other hand, the glass state in the present paper
refers to $T \le T_{\rm c}$.
Since $T_{\rm c}$ is located above $T_{\rm g}$, 
our theoretical results for glasses
apply also for some ranges of the liquid state in the 
conventional terminology.

According to our theoretical results, it is essential
to have the positive dispersion of the transversal 
sound velocity for the appearance of the low-frequency
excitations in the transversal current spectra,
whose presence has never been discussed so far. 
This comes out from our study on the density dependence
of the dispersion relation, shown in 
Figs.~\ref{fig:dispersion-T050-rho-1093}(a)
and \ref{fig:dispersion-T050-rho-950}(a)
for the higher and the lower densities,
respectively.
Such density dependence of the dispersion relation
is in accord with the one reported
from the computer simulation for vitreous silica
({\em cf.} Fig.~4 of Ref.~\cite{Pilla04}).
It would therefore be of value if simulation studies
could test whether or not our theoretical prediction
agrees with their data. 

We notice that the intensity
of the low-frequency excitations 
for the LJ system studied in this paper is weaker than the 
one, e.g., known from computer-simulation results
on silica~\cite{Horbach01b,Pilla04}:
the low-frequency excitations in the present study
are shoulders rather than peaks.
Again, this feature may be ascribed to the fact that the LJ system
can be classified as a fragile glass former, 
in which the boson peak is less
pronounced than in strong glass formers like
silica~\cite{Sokolov93b}.

\begin{figure}[tb]
\includegraphics[width=0.8\linewidth]{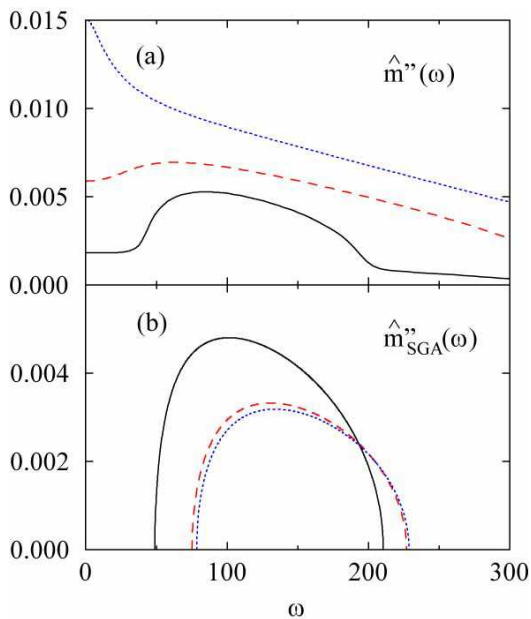}
\caption{
(a) The memory-kernel spectrum $\hat{m}''(\omega)$
and (b) the SGA spectrum $\hat{m}_{\rm SGA}''(\omega)$ 
obtained with the cutoff $q^{*} = 40$ (solid lines),
80 (dashed lines), and 120 (dotted lines)
for the hard-sphere system at packing fraction $\varphi = 0.6$.
As in Ref.~\cite{Goetze00}, 
the static structure factor $S_{q}$ 
has been evaluated within the
Percus-Yevick approximation, 
and the units of length and time have been chosen 
so that the hard-sphere diameter $d = 1$ and the
thermal velocity $v = 2.5$.}
\label{fig:HSS-cutoff}
\end{figure}

Finally, let us make a comment on the cutoff problem
mentioned in Ref.~\cite{Goetze00}.
As already noticed there, 
the MCT results for the hard-sphere system 
in stiff-glass states change if the cutoff $q^{*}$ 
of the wave-number grids used in the
calculations is varied.
This is demonstrated in Fig.~\ref{fig:HSS-cutoff}(a)
showing the $q^{*}$ dependence of the memory-kernel 
spectrum $\hat{m}''(\omega)$
for the hard-sphere system at packing fraction
$\varphi = 0.6$ studied in Ref.~\cite{Goetze00}.
It is seen that the intensity of $\hat{m}''(\omega)$
increases with increasing $q^{*}$. 
Furthermore, the broad harmonic-oscillation ``peak'' reflecting the AOP,
observable in $\hat{m}''(\omega)$ 
for the cutoff $q^{*} = 40$ adopted in Ref.~\cite{Goetze00},
disappears for $q^{*} = 120$;
it shows up only as a shoulder. 
Such $q^{*}$ dependence reflects the slow decrease towards zero
of the direct correlation function $c_{k}$ of the
hard-sphere system for $k$ tending to infinity,
because of which 
the relevant coupling coefficient $V_{k}$ 
({\em cf.} Eq.~(\ref{eq:MCT-phi-c}))
does not approach zero even in the $k \to \infty$ limit.
However, this does not mean that there is no AOP in the
hard-sphere system.
As discussed in Sec.~\ref{sec:sound-3}, the AOP portion
of $\hat{m}''(\omega)$, 
even when it is buried under the tail of the 
quasielastic peak, is well extracted by the SGA spectrum
$\hat{m}_{\rm SGA}''(\omega)$. 
Figure~\ref{fig:HSS-cutoff}(b) demonstrates that 
there is no $q^{*}$ dependence in $\hat{m}_{\rm SGA}''(\omega)$
provided a sufficiently large $q^{*}$ is chosen.
In this sense, the AOP is well defined in the hard-sphere system.
But, it does not show up
as a peak in $\hat{m}''(\omega)$ due to the
two-mode contributions, which are neglected within the SGA.
Thus, the increase of the intensity of $\hat{m}''(\omega)$,
in particular, the smearing out of the ``gap'' 
for $0 < \omega < \Omega_{-}$, with increasing $q^{*}$
is due to the development of the two-mode contributions.
Because of the strong anharmonic effects,
no harmonic-oscillation peak in the spectrum $\hat{m}''(\omega)$
is expected for the hard-sphere system.

On the other hand, 
introducing the cutoff is equivalent to 
softening the hard-sphere potential.
Indeed, we have confirmed that there is no
cutoff problem in the LJ system considered
in this paper,
and that all the essential results presented in Ref.~\cite{Goetze00}
can be reproduced for this system. 
In particular, one observes harmonic-oscillation peak
for the LJ system in stiff-glass states as shown in
Fig.~\ref{fig:hat-memories-T050}(a),
which is free from the mentioned cutoff problem. 
Thus, the whole physics discussed in Ref.~\cite{Goetze00}
for stiff-glass states remain valid for 
systems where particles interact with
regular interaction potentials.

\begin{acknowledgments}

The author thanks W.~G\"otze for 
valuable comments. 
This work was supported 
by Grant-in-Aids for scientific 
research from the 
Ministry of Education, Culture, Sports, Science and 
Technology of Japan (No. 17740282).

\end{acknowledgments}


\begin{thebibliography}{44}
\bibitem[{com({\natexlab{c}})}]{comment-Thomas-SGA}
\bibinfo{note}{The formulas~(\ref{eq:harmonic-app}) can be applied only when
  $w_{1} \le 1$ as discussed in Ref.~\cite{Goetze00}. We found, however, that
  $w_{1}$ exceeds unity for $\rho = 0.95$ and $T = 0.5$. Therefore, the SGA
  kernel $\hat{m}_{\rm SGA}(\omega)$ for this state has been obtained after
  solving the MCT equations (\ref{eq:GLE-hat-phi-d}) and
  (\ref{eq:MCT-hat-phi-a}) but dropping the two-mode contribution $\hat{\cal
  F}^{(2)}$ in the latter. As shown in Ref.~\cite{Thomas02}, the difference
  between the SGA kernel so obtained and the one based on
  Eqs.~(\ref{eq:harmonic-app}) is small. The threshold frequencies
  $\Omega_{\pm}$ for $\rho = 0.95$ and $T = 0.5$ have been determined as the
  frequencies where the intensity of $\hat{m}_{\rm SGA}(\omega)$ becomes 1\% of
  its peak height.}
\end{thebibliography}
\end{document}